\documentclass[a4paper,fleqn]{cas-dc}

\usepackage{natbib} 
\usepackage{notoccite}

\usepackage{graphicx}
\usepackage{multirow}
\usepackage{algorithm}
\usepackage{algpseudocode}
\usepackage{comment}
\usepackage{amsmath}
\usepackage{xcolor}
\hypersetup{colorlinks = true, allcolors=[RGB]{0,127,172}}
\usepackage{soul} 
\usepackage{array}
\usepackage{tabularray}
\usepackage{float}

\def\tsc#1{\csdef{#1}{\textsc{\lowercase{#1}}\xspace}}
\tsc{WGM}
\tsc{QE}
\tsc{EP}
\tsc{PMS}
\tsc{BEC}
\tsc{DE}

\begin{document}
\let\WriteBookmarks\relax
\def\floatpagepagefraction{1}
\def\textpagefraction{.001}

\shorttitle{Beneath the Surface}

\shortauthors{Ali Awad et~al.}

\title [mode = title]{Beneath the Surface: The Role of Underwater Image Enhancement in Object Detection}                      

\tnotetext[1]{This document is the results of the research
   project funded by the United States Geological Survey (USGS) under grant number G23AS00029.}


%
\author[1]{Ali Awad}

\cormark[1]


\ead{aawad@mtu.edu}


\credit{Conceptualization of this study, methodology, implementation, evaluation, and writing}

\affiliation[1]{organization={Department of Applied Computing, College of Computing, Michigan Technological University},
    addressline={1400 Townsend Dr}, 
    city={Houghton},
    postcode={49931}, 
    country={United States}}

\author[1]{Ashraf Saleem}

\credit{Conceptualization, methodology, validation, writing and supervision}

\author[2]{Sidike Paheding}[%
   ]

\credit{Methodology, analysis, writing and review}

\affiliation[2]{organization={Department of Computer Science and Engineering, Fairfield University},
    city={Fairfield},
    postcode={06824}, 
    state={CT},
    country={United States}}

\author[3]{Evan Lucas}[%
   ]

\credit{Methodology, data visualization, Writing and review}

\affiliation[3]{organization={Institute of Computing and Cybersystems, Michigan Technological University},
    city={Houghton},
    postcode={49931}, 
    state={Michigan},
    country={United States}}

\author[4]{Serein Al-Ratrout}[%
   ]

\credit{Literature survey}

\affiliation[4]{organization={Department of Computer Science, Michigan Technological University},
    city={Houghton},
    postcode={49931}, 
    state={Michigan},
    country={United States}}

\author[4]{Timothy C.~Havens}[%
   ]

\credit{Project administration, funding acquisition, and review}


\cortext[cor1]{Corresponding author}

\fntext[fn1]{Email address: aawad@mtu.edu (Ali Awad).}


\begin{abstract}
Underwater imagery often suffers from severe degradation resulting in low visual quality and reduced object detection performance. This work aims to evaluate state-of-the-art image enhancement models, investigate their effects on underwater object detection, and explore their potential to improve detection performance. To this end, we apply nine recent underwater image enhancement models, covering physical, non-physical and learning-based categories, to two recent underwater image datasets. Following this, we conduct joint qualitative and quantitative analyses on the original and enhanced images, revealing the discrepancy between the two analyses, and analyzing changes in the quality distribution of the images after enhancement. We then train three recent object detection models on the original datasets, selecting the best-performing detector for further analysis. This detector is subsequently re-trained on the enhanced datasets to evaluate changes in detection performance, highlighting the adverse effect of enhancement on detection performance at the dataset level. Next, we perform a correlation study to examine the relationship between various enhancement metrics and the mean Average Precision (mAP). Finally, we conduct an image-level analysis that reveals images of improved detection performance after enhancement. The findings of this study demonstrate the potential of image enhancement to improve detection performance and provide valuable insights for researchers to further explore the effects of enhancement on detection at the individual image level, rather than at the dataset level. This could enable the selective application of enhancement for improved detection. The data generated, code developed, and supplementary materials are publicly available at: \url{https://github.com/RSSL-MTU/Enhancement-Detection-Analysis}.
\end{abstract}




\maketitle

\section{Introduction}

Underwater computer vision has gained much attention in recent years with an increasing number of applications in areas such as ocean mapping, ecological monitoring, and inspection. This growth is largely attributed to the recent availability of large underwater datasets facilitated by the widespread use of Remotely Operated Vehicles (ROVs). Most recent datasets, such as \cite{jiang2021underwater, liu2021dataset, liu2021new}, are fully or partially obtained by ROVs due to their relatively low cost, availability, and ease of use. Underwater images are usually associated with limited visibility, reduced contrast, monochromatic casts, and low overall quality. This is attributed to two main phenomena: light absorption, by which the energy of the light is dissipated, and light scattering, by which the direction of light is changed \cite{yuan2022survey}.

Because of these challenges with underwater imagery, researchers have developed numerous enhancement approaches to mitigate these effects \cite{ACDC, zhang2023underwater, tang2022autoenhancer}. Assuming that image enhancement has a positive impact on high-level tasks, some studies \cite{li2016underwater, li2018benchmarking} suggested using image enhancement as a pre-processing step to increase classification and detection performance. However, other researchers concluded that image enhancement, unintuitively, degrades the detection performance. For example, Pei et al.~\cite{pei2019effects} concluded that image degradation negatively impacts classification performance but image enhancement algorithms do not help regain classification performance. The authors of \cite{fu2023rethinking, wang2023underwater} concluded that image enhancement has a negative effect on detection performance when used as a pre-processing step. Conversely, Chen et al.~\cite{chen2020reveal} concluded that mixing enhanced and non-enhanced images to train a classifier does not lead to higher performance, but rather to more robust performance.

Studies that address the use of Underwater Image Enhancement (UIE) for high-level tasks such as Underwater Object Detection (UOD) and classification often have limitations. For example, previous studies analyzed the effect of enhancement on entire object detection datasets. In contrast, this study involves qualitative and quantitative analyses at the individual image level. This is done by visually inspecting hundreds of images and generating per-image mAP for original and enhanced images. As a result, we reveal that the conclusions previously drawn on entire datasets do not always hold true for individual images, i.e., some images perform better after enhancement. This showcases the potential of enhancement in improving object detection. In addition, previous studies consider the quantitative and qualitative analyses separately. In this study, we generate figures to consider both types of analyses jointly, i.e., numerical values of quality are overlaid on the enhanced images and the detection results. Furthermore, we present a unique quality distribution analysis showing how enhancement degrades the quality of high-quality images.

This work conducts extensive experiments and analyses using nine enhancement algorithms, three detection algorithms, two datasets, and four image quality metrics to highlight the role of underwater image enhancement in object detection. The contributions of this work are listed below:
\begin{itemize}

\item We provide a joint quantitative and qualitative analysis, where images and their corresponding quality metric values are evaluated simultaneously. This offers insights on the reliability of existing underwater image quality metrics. Furthermore, this analysis is further extended to include detection results, enabling a joint evaluation of enhancement and detection.
\item We present a unique quality distribution analysis that illustrates how enhancement affects images of varying quality levels.
\item We study the relationship between image quality metrics and the mean Average Precision (mAP), investigating whether image quality could predict the detection performance.
\item We conduct a thorough image-level analysis instead of a dataset-level analysis, revealing the potential of image enhancement to improve object detection performance.

\end{itemize}

The remainder of this work is organized as follows. Section \ref{sec:related} surveys recent enhancement-detection studies, state-of-the-art (SOTA) underwater image enhancement models, and recent object detection models. Details of the experimental setup are provided in Section \ref{sec:eval}. Section \ref{sec:enhancement} analyzes the obtained image enhancement results, whereas Section \ref{sec:detection} analyzes the obtained object detection results. Then we analyze and discuss the effect of image enhancement on detection performance in Section \ref{sec:discussion}. Finally, we conclude the paper in Section \ref{sec:conclusion}.

\section{Literature Survey}
\label{sec:related}
In this section, we review combined enhancement-detection studies, Underwater Image Enhancement (UIE) models, and Underwater Object Detection (UOD) models.

\subsection{Combined Enhancement-Detection Studies}
A number of studies addressed the effect of image enhancement on object detection. For instance, the authors of \cite{chen2020reveal} studied the effect of mixed domains and different environments on detection and discussed the enhancement role in the process. The authors concluded that image restoration does not increase within-domain detection performance, but rather produces a more generalizable performance. One of the limitations of the study presented in \cite{chen2020reveal} is that it is limited to a single restoration technique. In a related effort, Pei et al.~\cite{pei2019effects} examined the effect of synthetic image degradation and restoration on classification performance. This study concluded that although image degradation severely affects the classification performance, image enhancement (i.e., degradation removal) does not lead to a restored or improved performance. This study can be extended to other higher-level tasks such as object detection. In another study, \cite{alawode2023improving} introduced a new video dataset called Underwater Visual Object Tracking (UVOT) and an enhancement model designed to improve object tracking performance. The study claimed significant detection performance improvement. However, no statistical significance analysis was provided in the study to validate the authors' claimed improvement. Another group of researchers \cite{wang2023underwater} further questioned the effect of enhancement on detection. The authors used the Toolbox for Identifying Object Detection Errors (TIDE) \cite{bolya2020tide} to conduct their analysis, which revealed that enhancement increases the False Positive (FP) rate because it changes the objects' edges that are essential for detection. In a recent development in image enhancement and object detection, Fu et al.~\cite{fu2023rethinking} proposed a new method for combining image enhancement and object detection using a shared loss function to train both the enhancer and detector simultaneously. The findings of their research indicate that the traditional enhancement approach of pre-processing the images with enhancement leads to a lower detection performance while a joint-learning approach increases it. The authors do not provide many details about the implementation of their proposed joint-learning framework. Moreover, the authors in \cite{wang2023reinforcement} introduced a novel enhancement-detection joint-learning framework based on reinforcement learning (RL). An RL agent is trained on the URPC2018 \cite{URPC} dataset with an action space consisting of a group of enhancement effects, a state space represented by image features, and detection performance as a reward. The authors claimed an improvement over the traditional pre-processing approach. This work could be expanded by extending the action space to include more enhancement effects and algorithms.

\subsection{Underwater Image Enhancement (UIE)}
Underwater image enhancement is categorized into three main groups \cite{xu2023systematic, zhou2023underwater, hu2022overview}: non-physical, physical, and learning-based methods. In general, non-physical models work directly on images using image processing techniques without using any prior knowledge, so they are fast and simple but often lead to sub-optimal enhancement details \cite{gu2023underwater}. For instance, Zhang et al.~\cite{ACDC} introduced a method for enhancing underwater images using a combination of various techniques. Their proposed method included attenuated color channel correction (ACCC), a fusion-based contrast improvement method, and multiscale unsharp masking (MSUM) strategy. Although this method generalizes well to low-light images and hazy images, it may reduce the color quality of already low-quality images. This method is also computationally expensive. The retinex-based enhancement model in \cite{zhuang2021bayesian} used multi-order gradient priors of reflectance and illumination. This approach is based on a simple and effective color correction approach to remove color casts and recover naturalness. Then a maximum posterior formulation is used by imposing multi-order gradient priors on both reflectance and illumination. Another model called Texture Enhancement Model based on Blurriness and Color Fusion (TEBCF) is presented in \cite{yuan2021tebcf}. This model used multi-scale fusion to merge two inputs; one to improve contrast based on dark channel prior in the RGB space and another to improve color based on the CIELAB color space \cite{chen2019towards}. Although the method introduced in \cite{yuan2021tebcf} showed a significant quantitative performance advantage over other approaches, it did not perform as good qualitatively.

In contrast to the non-physical methods described, physical methods incorporate some kind of prior knowledge based on the physics of the underwater image formation process. However, the assumptions made for unknown parameters may not be entirely suitable or precise in complex underwater environments, which restricts their use cases \cite{guo2017research}. For example, Zhang et al.~\cite{zhang2023underwater} developed an approach that begins with a color cast correction method for each color channel. Subsequently, various enhancement methods are applied to enhance the base and detail layers of the \emph{V} channel in the \emph{HSV} space, which is decomposed using the spatial prior and texture prior. This method showed good generalization capability for fog and low-light images. In another enhancement approach \cite{hou2023non}, a new model called Illumination Channel Sparsity Prior (ICSP) was developed based on the observation that there are some low-intensity pixels in the illumination (\emph{I}) channel of an underwater image with uniform lighting in the \emph{HSI} color space. Using this observation, the authors developed a variational model based on the retinex theory \cite{McCann2016}. However, some images produced by this model are over-brightened.

Unlike both physical and non-physical enhancement methods, learning-based enhancement methods rely on neural networks and can generate high-quality enhancement results but require parameter tuning and a sufficient amount of data \cite {liao2024underwater}, which can be challenging for underwater environments. Researchers in \cite{huang2023contrastive} introduced a Mean Teacher-based Semi-supervised Underwater Image Restoration (Semi-UIR) model which incorporates unlabeled data into network training. Similarly, Fu et al.~\cite{fu2022unsupervised} introduced an UnSupervised Underwater Image Restoration (USUIR) method by leveraging the homology property between a raw underwater image and a 're-degraded' image. This proposed approach decomposes the underwater image into three latent components. The design of this model allows it to run relatively fast and at the same time achieve good qualitative results. In \cite{wang2023domain}, a Two-phase Underwater Domain Adaptation network (TUDA) was introduced, which simultaneously minimizes the inter-domain and intra-domain gaps. First, a triple-alignment network was designed to jointly perform image-level, feature-level, and output-level alignment using adversarial learning to reduce the inter-domain gap. Then, an easy/hard adaptation technique was developed to reduce intra-domain gaps. The colors produced by this method are vibrant and the images are visually pleasing. Another group of researchers~\cite{tang2022autoenhancer} applied Neural Architectural search (NAS) to search for the optimal U-Net architecture specifically tailored for underwater image enhancement. This approach demonstrated good visual performance across different underwater scenarios. A different approach was explored in \cite{sun2022underwater}, where underwater image enhancement was modeled as a Markov decision process (MDP). This approach was based on reinforcement learning frameworks that selected a set of image enhancement actions and organized them into an optimal sequence. This is one of the preliminary studies on the use of reinforcement learning in underwater image enhancement.

\subsection{Underwater Object Detection (UOD)}
Object detection algorithms can be roughly categorized into one-stage and two-stage detectors. One-stage detectors generate object coordinates and labels in a single step, making them faster but generally less accurate. In contrast, two-stage detectors use a separate region proposal network to first identify potential object locations before classifying the objects, resulting in higher performance at the cost of speed. A group of researchers developed two-stage detectors for underwater scenarios. For example, Mandal et al.~\cite{mandal2018assessing} integrated the Faster R-CNN \cite{ren2016faster} with three distinct backbone networks to detect fish species. In another work, Lin et al.~\cite{lin2020roimix} introduced a data augmentation technique that used candidate region fusion to create training samples that mimic overlap, occlusion, and blurring effects. Xu et al.~\cite{xu2021scale} developed a scale-aware feature pyramid detector based on a unique backbone sub-network which captures the fine-grained features of smaller targets. Qi et al.~\cite{qi2022underwater} introduced a detection model based on a deformable convolutional pyramid structure to identify small underwater objects. Song et al. \cite{song2023boosting} proposed a new model comprised of a region proposal network that provided the prior probability of objects, then the classification score and the prior uncertainty were combined to generate the score of the final prediction. Finally, the classification loss is increased for miscalculated proposals, whereas it is reduced for accurately predicted proposals.

Other works have focused on one-stage underwater detectors. For instance, Sung et al.~\cite{sung2017vision} focused on detecting fish in real-time using a YOLO-v1 \cite{redmon2016you}-based detection framework. Similarly, Hu et al.~\cite{hu2020marine} developed a detection network for sea urchins based on the detection algorithm presented in \cite{liu2016ssd}. A novel multi-directional edge detection technique was introduced to better capture the distinctive spiny edge characteristics of sea urchins, thereby improving feature representation. Chen et al.~\cite{chen2020underwater} introduced a neural network architecture called SWIPENet based on the framework presented in \cite{fu2017dssd}. This model was designed for the detection of small targets in underwater environments. Furthermore, the YOLO architecture has been used in several works. For instance, Zhang et al.~\cite{zhang2021lightweight} introduced a lightweight approach for recognizing underwater objects based on YOLO-v4 \cite{bochkovskiy2020yolov4}. Liu et al.~\cite{liu2023underwater} proposed TC-YOLO, a novel underwater object detection technique that includes attention mechanisms by inetgrating a coordinate attention module and a transformer encoder into YOLO-v5 \cite{ultralytics2021yolov5}. YOLO-NAS \cite{supergradients} introduced a novel algorithmic optimization engine to the YOLO family known as \emph{Automated Neural Architecture Construction} (AutoNAC) which guarantees optimal hardware utilization while maintaining baseline performance.

\section{Experimental Setup}
\label{sec:eval}
To support our study, we selected representative image enhancement models covering the main image enhancement categories. The selected non-physical methods include ACDC \cite{ACDC}, TEBCF \cite{yuan2021tebcf}, BayesRet ~\cite{zhuang2021bayesian}, covering histogram-based, fusion-based, and retinex-based enhancement. The selected physical methods include PCDE \cite{zhang2023underwater} and ICSP \cite{hou2023non}, covering dark channel prior and optical imaging properties-based enhancement. The selected deep learning methods include AutoEnh \cite{tang2022autoenhancer}, Semi-UIR \cite{huang2023contrastive}, USUIR \cite{fu2022unsupervised}, and TUDA \cite{wang2023domain} covering transformer-based, contrastive-learning-based, and GAN-based models. The source codes and instructions to run these enhancement models can be found in a GitHub link within their corresponding provided references. We implemented all enhancement models without modification or additional training, using the source codes and the trained models provided by the original authors.

On the detection side, we selected four common object detection algorithms, including YOLO-NAS \cite{supergradients}, RetinaNet \cite{wu2019detectron2}, and Faster R-CNN \cite{wu2019detectron2} covering one-stage and two-stage detectors. The top-performing detector on the original images was selected for the subsequent evaluations. This includes training separate models on the enhanced images from each image enhancement algorithm for both datasets totaling 20 model instances, each trained and evaluated separately. The image resolution for both datasets is set to $800\times 600$. We used the SuperGradients \cite{supergradients} library to implement the detector on a Linux server with two Nvidia Tesla V100 GPUs. The training was performed with a \emph{batch size} of 16 and the AdamW optimizer with a \emph{weight decay} of 0.00001. We use the YOLO-NAS Large (L) architecture and start from the COCO pre-trained weights.

As for the selected datasets, we use two publicly available datasets to ensure the generalizability of our conclusions, namely CUPDD \cite{saleem2023multi} and RUOD \cite{fu2023rethinking}. The RUOD is one of the largest publicly available underwater datasets, combining images from different datasets and containing different color casts, domains, environments, and aquatic creatures. It is comprised of 14,000 high-resolution underwater images, 9,800 of which are used for training, with approximately 75,000 total box annotations and 10 different classes. On the other hand, the CUPDD is a relatively smaller dataset containing severely degraded challenging images. It comprises 414 images, 313 of which are used for training, with three general categories of aquatic plants collected in the Great Lakes region in the US. The use of both datasets helps us draw generalized conclusions because the RUOD is very large, recent, and extensive. In contrast, CUPDD is very challenging and has the potential to test enhancement algorithms at the extreme end. 

We opt for a reference-free image quality evaluation because we do not have reference images (i.e., ground truth images with no water effects) using four metrics: 1) the Underwater Image Quality Measure (UIQM) \cite{panetta2015human}, which comprises color, contrast, and sharpness indices. A higher UIQM score indicates better image quality. 2) The Underwater Color Image Quality Evaluation (UCIQE) \cite{yang2015underwater}, which represents a linear combination of saturation, contrast and chroma. A higher UCIQE score indicates better image quality. 3) An underwater metric based on light absorption and scattering characteristics (CCF) \cite{wang2018imaging},  which is a linear regression model based on colorfulness, contrast, and fog density indices. A higher CCF score indicates better image quality. 4) The Matlab implementation of Entropy \cite{gonzalez2003digital}. A higher Entropy score indicates richer details, i.e., higher image quality. All of the metrics used were applied without any modification and are implemented with original weights and hyperparameter values from the original papers.

To facilitate and focus our image quality distribution analysis, we combine the four selected IQMs into a single quality index (Q-index). The Q-index is used throughout this work as a general indication of image quality and is used to describe and analyze the quality distribution of images before and after enhancement. Essentially, the Q-index is generated from the four selected metrics through outlier removal, global re-scaling, and averaging processes to produce a single bounded metric between zero and one. A value is flagged as an outlier if it is three Median Absolute Deviations (MADs) away from the median based on the MATLAB implementation in \cite{mathworksDetectReplace}. Moreover, global re-scaling means that the minimum and maximum values are taken across all images from the original and enhanced datasets to fairly compare the Q-index values across different models.

\begin{table}[t]
\centering
\caption{Quantitative evaluation of the selected enhancement models on the \textbf{CUPDD} \cite{saleem2023multi} dataset using the mean and standard deviation (std) values of the UIQM \cite{panetta2015human}, UCIQE \cite{yang2015underwater}, CCF \cite{wang2018imaging} and Entropy \cite{gonzalez2003digital}. Highest scores are highlighted in bold.}
\label{tab:quant-CUPDD}
\resizebox{\columnwidth}{!}{%
\begin{tabular}{lcccccccc}
\hline

\multicolumn{1}{c|}{\multirow{2}{*}{\textbf{Models}}} &
  \multicolumn{2}{c}{\textbf{UIQM ↑}} &
  \multicolumn{2}{c}{\textbf{UCIQE ↑}} &
  \multicolumn{2}{c}{\textbf{CCF ↑}} &
  \multicolumn{2}{c}{\textbf{Entropy ↑}} \\ \cline{2-9} 
\multicolumn{1}{c|}{}          & Mean & Std. & Mean & Std. & Mean  & Std. & Mean & Std. \\ \hline
\multicolumn{1}{l|}{Original}  & 1.11 & 0.35 & 0.46 & 0.03 & 14.52 & 4.23 & 7.21 & 0.28 \\
\multicolumn{1}{l|}{ACDC \cite{ACDC}}      & 3.07 & 0.54 & 0.51 & 0.02 & 14.97 & 2.93 & 7.52 & 0.11 \\
\multicolumn{1}{l|}{TEBCF \cite{yuan2021tebcf}} &
  \textbf{4.07} &
  0.69 &
  \textbf{0.61} &
  0.02 &
  \textbf{25.03} &
  5.01 &
  \textbf{7.80} &
  0.09 \\
\multicolumn{1}{l|}{BayesRet \cite{zhuang2021bayesian}} & 2.88 & 0.66 & 0.55 & 0.03 & 13.17 & 3.86 & 7.72 & 0.05 \\
\multicolumn{1}{l|}{PCDE \cite{zhang2023underwater}}      & 1.27 & 0.31 & 0.46 & 0.02 & 7.55  & 1.42 & 6.93 & 0.28 \\
\multicolumn{1}{l|}{ICSP \cite{hou2023non}
}      & 1.17 & 0.40 & 0.43 & 0.04 & 10.57 & 3.04 & 4.56 & 0.96 \\
\multicolumn{1}{l|}{AutoEnh \cite{tang2022autoenhancer}}  & 1.98 & 0.57 & 0.54 & 0.03 & 10.79 & 2.53 & 7.55 & 0.19 \\
\multicolumn{1}{l|}{Semi UIR \cite{huang2023contrastive}}  & 3.21 & 0.59 & 0.55 & 0.04 & 13.42 & 4.12 & 7.56 & 0.19 \\
\multicolumn{1}{l|}{USUIR \cite{fu2022unsupervised}}     & 1.65 & 0.42 & 0.60 & 0.02 & 19.13 & 4.35 & 7.67 & 0.15 \\
\multicolumn{1}{l|}{TUDA \cite{wang2023domain}}      & 2.58 & 0.56 & 0.59 & 0.02 & 12.62 & 2.33 & 7.77 & 0.07 \\ \hline
\end{tabular}%
}
\end{table}

\section{Enhancement Evaluation}
\label{sec:enhancement}
In this section, a quantitative and qualitative evaluation of the selected image enhancement models is presented for both the RUOD and CUPDD datasets. In addition, we generate the Q-index based on the selected metrics to show the quality distribution of the images after enhancement and to provide a joint quantitative and qualitative analysis.

\subsection{Quantitative Enhancement Evaluation}
The images from the selected datasets are enhanced using the selected enhancement models and evaluated using the selected image quality metrics. On the one hand, the results for the CUPDD dataset are shown in Table \ref{tab:quant-CUPDD}. These results show the high performance of the TEBCF model \cite{yuan2021tebcf} based on all metrics, especially in terms of the UIQM valued at 4.07 and CCF valued at 25.03. This should indicate its capability of producing vibrant colors, marked contrast, soft edges, and removing fog effects. The Semi-UIR model also performs notably with a UIQM of 3.21 and high UCIQE and Entropy values (0.55 and 7.56, respectively), suggesting that it offers a balanced image quality enhancement with rich details. Meanwhile, the USUIR model stands out in UCIQE (0.60) and Entropy (7.67), reflecting its proficiency in improving colorfulness and retaining image detail. The ACDC is another notable model that signifies a marked enhancement capability with a UIQM of 3.07 and balanced performance on other metrics. Similarly, BayesRet and TUDA demonstrate a balanced approach with a commendable Entropy at 7.72 and 7.77, respectively, indicating substantial detail retention alongside reasonable color and contrast enhancements. Conversely, models like ICSP and PCDE, with lower UIQM and CCF values, reflect limited capabilities in quality enhancement and color correction. These comparative insights highlight the strengths and potential applications of each model in underwater image processing, emphasizing the importance of selecting a model based on the desired balance of quality, color correction, and detail retention. It is worth noting that all models are performing better than the Original except for the ICSP indicating that it is possibly distorting colors and introducing noise. The PCDE and the BayesRet are also less effective to the Original indicating their limited capabilities of enhancing underwater images.

\begin{table}[t]
\centering
\caption{Quantitative evaluation of the selected enhancement models on the \textbf{RUOD} \cite{fu2023rethinking} dataset using the mean and standard deviation (std) values of the UIQM \cite{panetta2015human}, UCIQE \cite{yang2015underwater}, CCF \cite{wang2018imaging} and Entropy \cite{gonzalez2003digital}. Highest scores are highlighted in bold.}
\label{tab:quant-RUOD}
\resizebox{\columnwidth}{!}{%
\begin{tabular}{lcccccccc}
\hline

\multicolumn{1}{c|}{\multirow{2}{*}{\textbf{Models}}} &
  \multicolumn{2}{c}{\textbf{UIQM ↑}} &
  \multicolumn{2}{c}{\textbf{UCIQE ↑}} &
  \multicolumn{2}{c}{\textbf{CCF ↑}} &
  \multicolumn{2}{c}{\textbf{Entropy ↑}} \\ \cline{2-9} 
\multicolumn{1}{c|}{}          & Mean          & Std. & Mean          & Std. & Mean           & Std. & Mean          & Std. \\ \hline
\multicolumn{1}{l|}{Original}  & 1.14          & 1.18 & 0.52          & 0.05 & 20.79          & 6.56 & 7.20          & 0.36 \\
\multicolumn{1}{l|}{ACDC \cite{ACDC}}      & 3.76          & 0.76 & 0.55          & 0.03 & 25.49          & 5.02 & 7.67          & 0.16 \\
\multicolumn{1}{l|}{TEBCF \cite{yuan2021tebcf}}     & 3.65          & 0.94 & \textbf{0.62} & 0.03 & \textbf{31.50} & 4.29 & 7.62          & 0.23 \\
\multicolumn{1}{l|}{BayesRet \cite{zhuang2021bayesian}} & \textbf{3.85} & 0.80 & 0.58          & 0.03 & 26.80          & 7.15 & \textbf{7.74} & 0.12 \\
\multicolumn{1}{l|}{PCDE \cite{zhang2023underwater}}      & 2.42          & 0.75 & 0.51          & 0.04 & 14.76          & 3.72 & 6.75          & 0.43 \\
\multicolumn{1}{l|}{ICSP \cite{hou2023non}
}      & 1.14          & 1.41 & 0.54          & 0.05 & 25.59          & 7.87 & 6.78          & 0.77 \\
\multicolumn{1}{l|}{AutoEnh \cite{tang2022autoenhancer}
}  & 2.78          & 1.17 & 0.59          & 0.04 & 22.05          & 5.93 & 7.48          & 0.31 \\
\multicolumn{1}{l|}{Semi UIR \cite{huang2023contrastive}}  & 3.07          & 1.17 & 0.60          & 0.04 & 26.84          & 8.36 & 7.60          & 0.22 \\
\multicolumn{1}{l|}{USUIR \cite{fu2022unsupervised}}     & 2.63          & 0.84 & \textbf{0.62} & 0.04 & 25.16          & 7.17 & 7.53          & 0.26 \\
\multicolumn{1}{l|}{TUDA \cite{wang2023domain}}      & 3.40          & 0.95 & 0.58          & 0.02 & 20.88          & 4.52 & 7.65          & 0.17 \\ \hline
\end{tabular}%
}
\end{table}

On the other hand, the results for the RUOD dataset are shown in Table \ref{tab:quant-RUOD}. These results show relatively higher CCF and Entropy scores compared to the results on the CUPDD dataset. This could be attributed to the fact that RUOD contains clearer images with richer details. In a similar performance jump, the TEBCF \cite{yuan2021tebcf} is a top-performer in terms of the UCIQE and the CCF at 0.62 and 31.5, respectively. The TEBCF is followed by the BayesRet at top UIQM and entropy values of 3.85 and 7.74, respectively. This might indicate the capability of the TEBCF in removing the fog effect and the capability of the BayesRet in adding more details. The ACDC also shows robust performance with a high UIQM (3.76) and balanced metrics, rendering it a versatile candidate for underwater image enhancement. TUDA further demonstrates solid capability with a UIQM of 3.40, a UCIQE of 0.58, a CCF of 20.88, and a high Entropy of 7.65, highlighting it as a strong performer overall, particularly excelling in maintaining detail. Models like AutoEnh and Semi-UIR are notable for their good colorfulness and contrast enhancement at a UCIQE of 0.59 and 0.60, respectively, with reasonable overall quality and detail retention. In contrast, PCDE and ICSP exhibit lower UIQM (2.42 and 1.14) and CCF (14.76 and 25.59), reflecting their relatively less impactful enhancements. Aligning with the results from Table \ref{tab:quant-CUPDD}, All models outperform the Original except for the ICSP and the PCDE on some metrics reflecting the limited capabilities of both models.

\begin{figure}[t]
\centering
\includegraphics[width=\columnwidth]{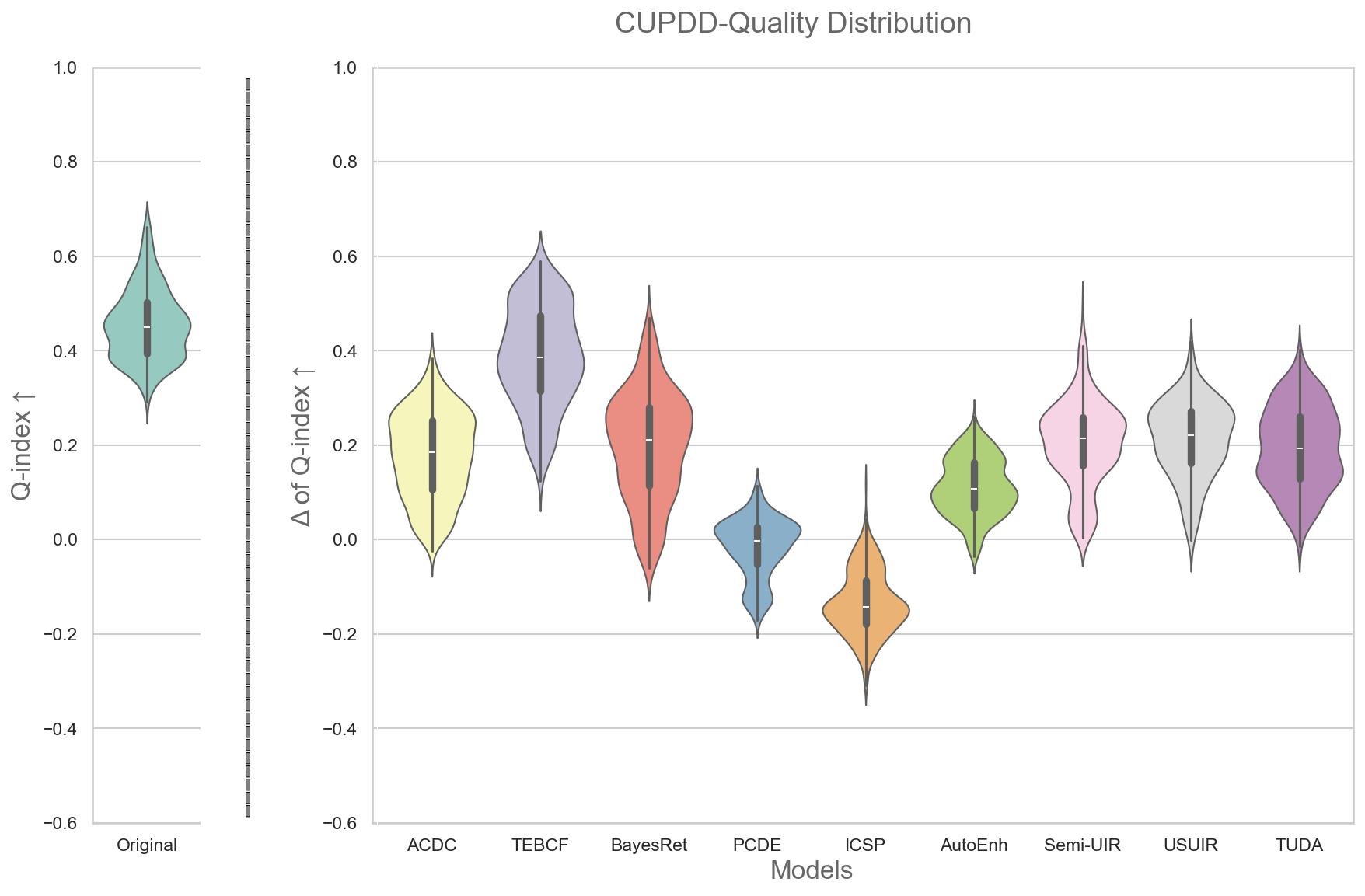}
  \caption{On the left, The quality distribution of the Original images of the CUPDD dataset \cite{saleem2023multi} based on the Q-index. On the right, the distributions of the change in quality after enhancement by different models.}
  \label{fig:Q-Dist-CUPDD}
\end{figure}

\subsection{Quality Distribution}
\label{q-dist}
\begin{figure}[t]
\centering
\includegraphics[width=\columnwidth]{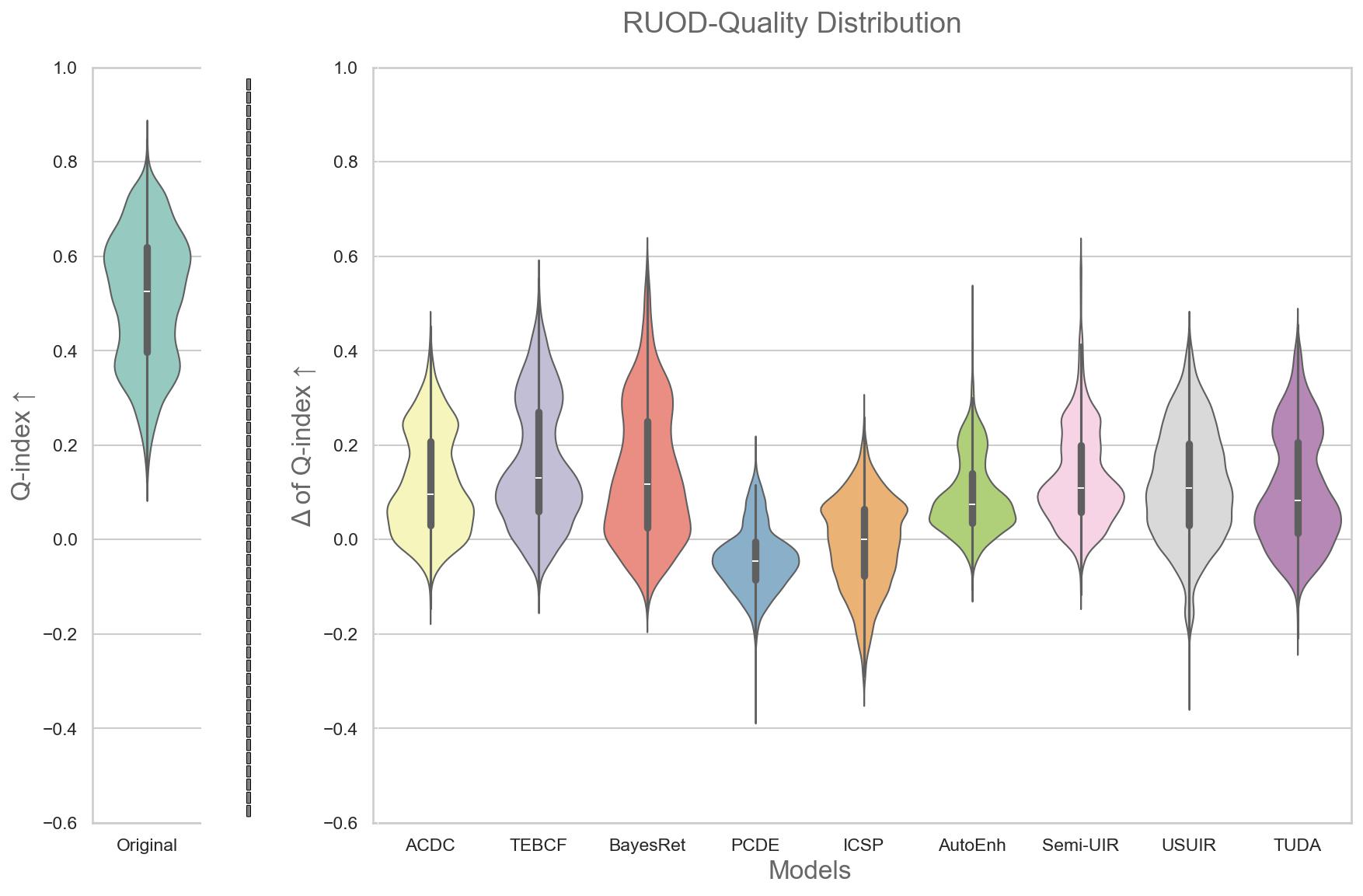}
  \caption{On the left, The quality distribution of the Original images of the RUOD dataset \cite{fu2023rethinking} based on the Q-index. On the right, the distributions of the change in quality after enhancement by different models.}
  \label{fig:Q-Dist-RUOD}
\end{figure}

We used the proposed Q-index in section~\ref{sec:eval} to facilitate an analysis of the quality distribution of images before and after enhancement by each enhancement model. In particular, a violin plot is used to show the quality distribution of original images and the change in the quality distribution after enhancement, as shown in Figs.~\ref{fig:Q-Dist-CUPDD} and \ref{fig:Q-Dist-RUOD}. The change in the Q-index ($\Delta$ Q-index) for an image is simply calculated as 
\begin{align}
\begin{split}
\Delta \; \; Q\textrm{-}index = Q\textrm{-}index_{(Enh)} - Q\textrm{-}index_{(Org)},
\end{split}
\end{align}
where $(Enh)$ refers to the enhanced image and $(Org)$ refers to the original. This helps us understand how enhancement is changing the quality distribution of the images, i.e., whether enhancement uniformly improves the quality of images across the quality spectrum. On the one hand, by examining Fig.~\ref{fig:Q-Dist-CUPDD}, it is observed that the entire original image dataset ranges from 0.25 to 0.75, with an almost bimodal distribution and a median of approximately 0.45. One peak is centered around the median, and the other at just below 0.4, indicating that most images lie around those two peaks. Moving to the distribution of images after enhancement, we notice that all models, except for the PCDE and ICSP, produced a positive distribution. The PCDE, with almost half the distribution in the negative, and ICSP, with the entire distribution in the negative, generally produce images that have lower quality than the originals. The TEBCF is a top-performer at a median of approximately 0.38, followed by the ACDC, BayesRet, Semi-UIR, USUIR, and TUDA at a median of approximately 0.2. In contrast, AutoEnh showed lower performance exhibiting a bimodal distribution with most of the images centered around 0.1. It is also noted that for ACDC, TEBCF, BayesRet, Semi-UIR, and USUIR, the distribution has peaks at the top and tapering off towards the bottom indicating that these models are shifting low-quality images into high-quality images. A small number of images performed worse after enhancement with the ACDC, TEBCF, AutoEnh, Semi-UIR, USUIR, and TUDA models.

On the other hand, the distribution of the original RUOD images shown in Fig. \ref{fig:Q-Dist-RUOD} is a clearer bimodal distribution with a median of nearly 0.52 and two distinct peaks at 0.6 and 0.35. Furthermore, most enhancement models have a median at 0.1 except for the PCDE and ICSP, which exhibit similar negative trends as in Fig. \ref{fig:Q-Dist-CUPDD}. A notable observation about the distribution of the enhanced images in Fig. \ref{fig:Q-Dist-RUOD} is that most models have an inverted distribution compared to the Original with a larger peak at the bottom. This means that enhancement is \emph{deteriorating the quality of high-quality images  by over-enhancing them} in the RUOD dataset. ON the contrary, CUPDD had the larger peak on top. This could be attributed to the fact that CUPDD images are mostly severely degraded images and have significantly lower quality than RUOD. This supports that \emph{enhancement is more beneficial to low-quality images}.
 
\begin{figure}[ht]
\centering
\includegraphics[width=\columnwidth]{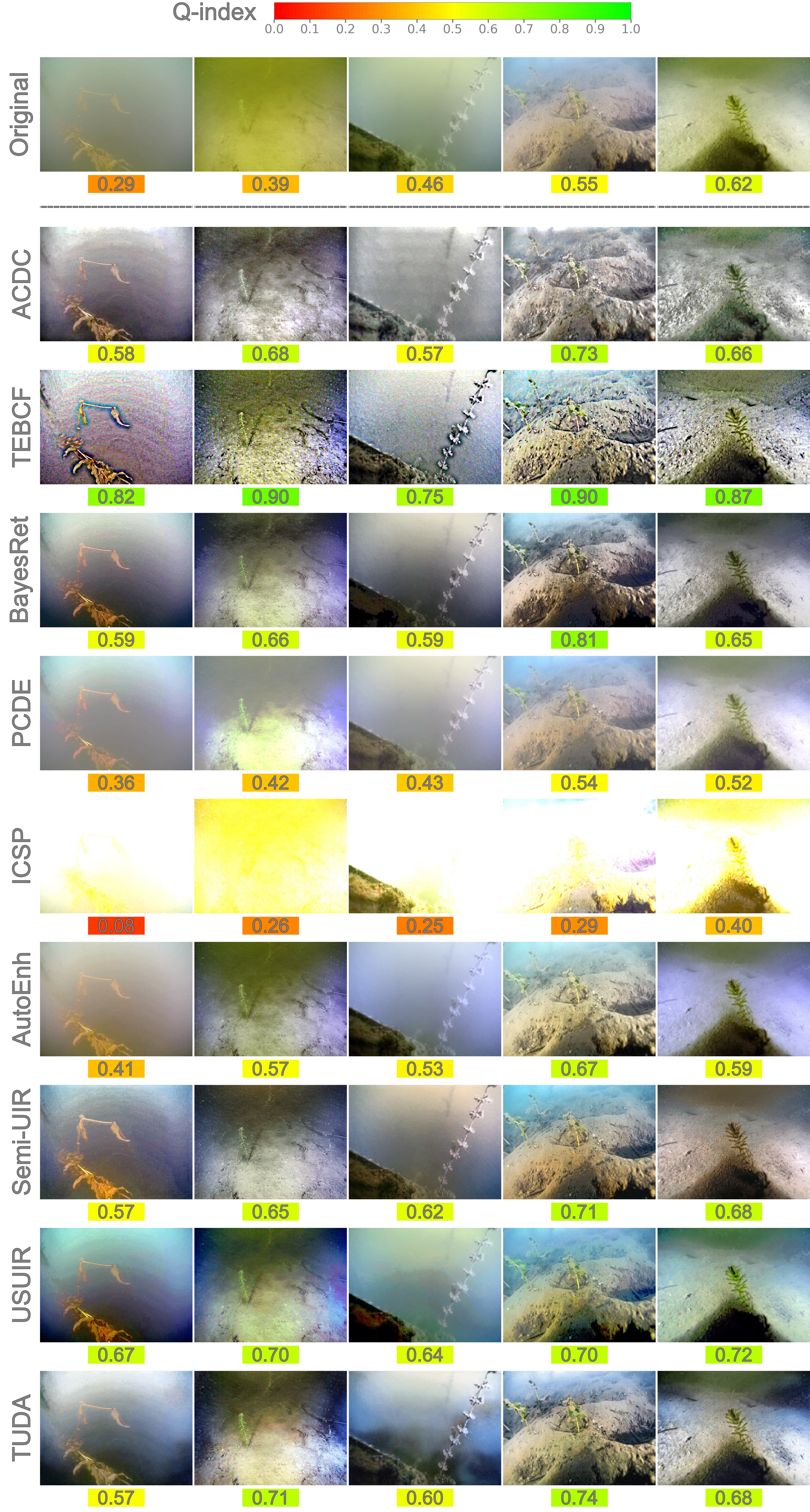}
  \caption{Randomly selected Original images from each available quality bin of CUPDD dataset. The corresponding Q-index values are color-mapped and placed under each image.}
  \label{fig:cuppd_qual}
\end{figure}

\begin{figure*}[ht]
\centering
\includegraphics[width=0.99\textwidth]{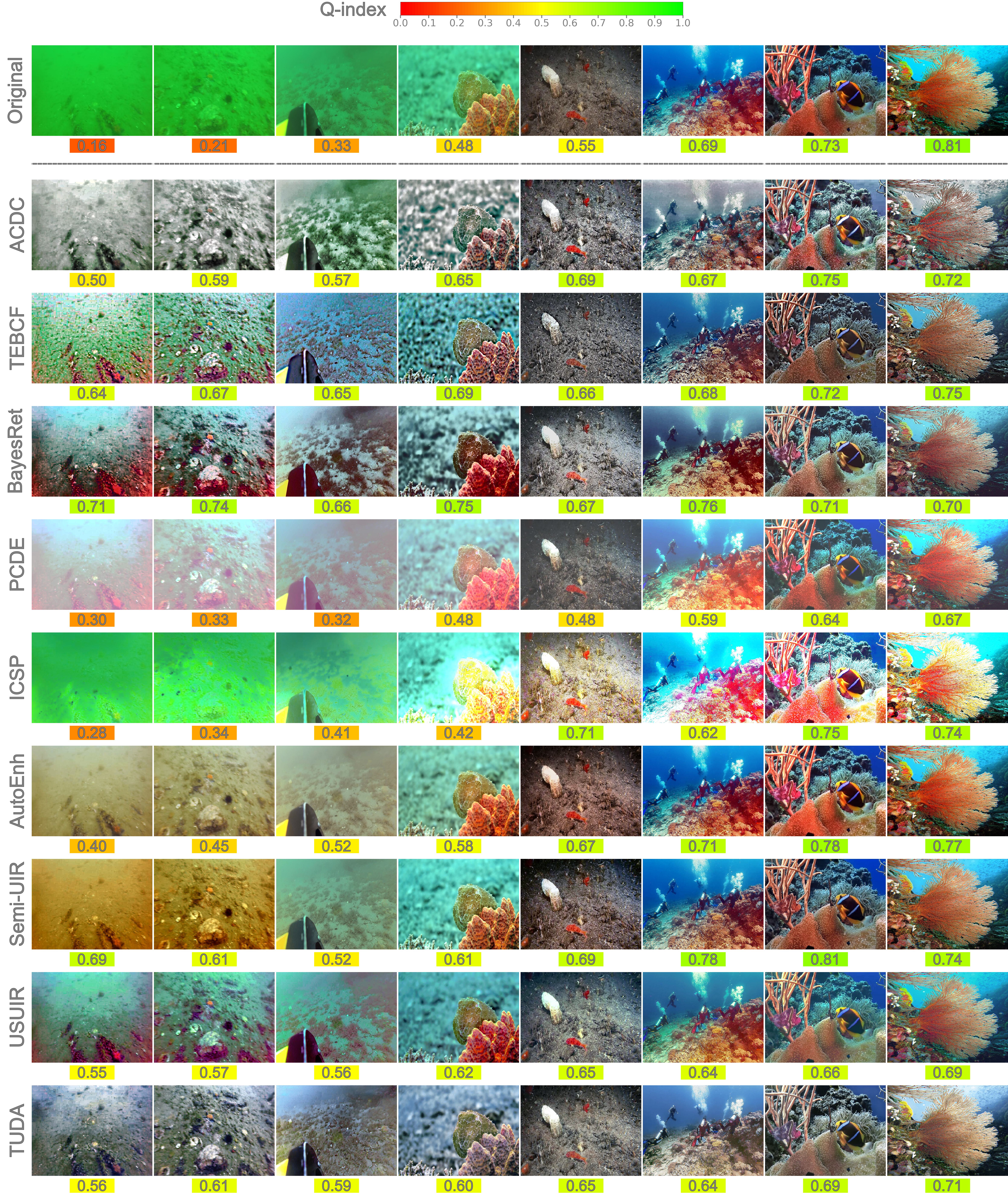}
  \caption{Randomly selected Original images from each available quality bin of RUOD dataset. The corresponding Q-index values are color-mapped and placed under each image.}
\label{fig:ruod_qual}
\end{figure*}

\subsection{Joint Quantitative-Qualitative Enhancement Evaluation}
As a further and final evaluation of image enhancement, we categorize images into 10 quality bins (levels) based on their Q-index values and randomly select an image representing each quality bin for both datasets, namely the CUPDD and the RUOD as shown in Figs. \ref{fig:cuppd_qual} and \ref{fig:ruod_qual}. Some quality bins at both ends are omitted since no images are found in those extreme bins. The Q-index value is shown under each image and color-mapped from red (low-quality) to green (high-quality) for easier perception of the quantitative quality. Fig  \ref{fig:cuppd_qual} shows a random image from each quality bin available in the CUPDD dataset with their enhanced versions and corresponding Q-index values. A gradual improvement in quality can be seen as the Q-index value of the Original images increases from left to right. However, we notice that the image belonging to the 0.5-0.6 quality bin looks more visually pleasing than its successor with softer details, less noise, and more vibrant colors. The successor original image at a higher Q-index of 0.62 displays to a strong color cast and more noise, yet achieves a higher Q-index, which questions the reliability of the current underwater image quality metrics. This observation is further demonstrated by the enhanced images, especially with the TEBCF images, where they all look very sharp and noisy but achieve Q-index values exceeding 0.75. In contrast, the Q-index aligns well with our qualitative evaluation of the PCDE and the ICSP models at values ranging from 0.52 to values as low as 0.08. The PCDE produces very foggy images whereas the ICSP produces extremely overexposed images, further deteriorating the quality of the  original images. This suggests that although current underwater metrics are not entirely robust, they can still provide a general indication of image quality. The ACDC and BayesRet produce less saturated color tones but maintain the details of the images. The AutoEnh, Semi-UIR, USUIR and the TUDA seem to be producing the most visually pleasing images with vibrant colors, less noise and fog effects compared to other models. However, this is not always reflected by their Q-index values. It is notable that the average increase in the Q-index value of low-quality images is much higher than the average increase in the Q-index values of high-quality images. For instance, the Q-index of the image from the lowest quality bin nearly doubled with some models whereas the Q-index of the image from the highest quality bin showed only a marginal improvement with most models. In fact, most high-quality original images in the RUOD Fig. \ref {fig:ruod_qual} attained lower Q-index values after enhancement. This supports the conclusion from our quality distribution analysis in section \ref{q-dist} that enhancement of high-quality images often results in over-enhancement.

In accordance with the CUPDD, the random images from the RUOD dataset shown in Fig. \ref{fig:ruod_qual} exhibit a similar gradual improvement from left to right. Images with low Q-index tend to have a monochromatic cast with very few details and low visibility. A less severe color cast with more details and colors can be noted as the Q-index increases. Images with high Q-index tend to have a richer red channel, more vibrant color, and clearer fine details. Although it is hard to distinguish a slight variation in the Q-index, the general trend of quality improvement with the increase of the Q-index is evident. For instance, the quality of the first two images at 0.16 and 0.21 is hardly distinguishable, but the difference between the first and last images is highly notable. The performance of the enhancers on the RUOD is very similar to CUPDD with TEBCF quantitatively leading but qualitatively lagging behind. This discrepancy in the quantitative and qualitative further highlights the limitation of current underwater metrics. We identify the TUDA and the Semi-UIR as the best enhancers for producing visually pleasing images. We conclude that enhancement techniques are most effective on low-quality images and less effective on high-quality images. Also, we observe that the selected deep learning-based enhancement models produce more visually pleasing images than traditional models. Lastly, current underwater image quality metrics can be used as a general indication of image quality, despite their limitations. 

\section{Detection Evaluation}
\label{sec:detection}
In this section, we first compare the detection performance of the selected three detection algorithms on the original images as shown in Table \ref{tab:all-DET}, where YOLO-NAS outperformed all other algorithms by a large margin, followed by RetinaNet and finally Faster R-CNN. Therefore, YOLO-NAS is selected as the best-performing detector and a total of 20 models are created from YOLO-NAS, including nine models for the enhanced versions of the CUPDD, nine models for the enhanced versions of the RUOD, and two models for the original images of the CUPDD and the RUOD. The YOLO detection family has been extensively used in underwater scenarios demonstrating its efficiency \cite{redmon2016you, zhang2021lightweight, liu2023underwater} with YOLO-NAS being one of the recent top performers \cite{hamzaoui2024efficient, ercan2023gesture, awad2024underwater}.

\begin{table}
\centering
\label{tab:all-DET}
\caption{Comparison of the Detection performance of different object detection algorithms on the original images of the RUOD dataset \cite{fu2023rethinking} using the mean Average Precision (mAP).}
\label{tab:CUPDD-DET}
\resizebox{\linewidth}{!}{%
\begin{tabular}{>{\hspace{0pt}}m{0.602\linewidth}|>{\centering\hspace{0pt}}m{0.135\linewidth}>{\centering\arraybackslash\hspace{0pt}}m{0.192\linewidth}} \hline
\multicolumn{1}{>{\centering\hspace{0pt}}m{0.602\linewidth}|}{Methods} & mAP$_{50}$~ & mAP$_{50-95}$~ \\ \hline
YOLO-NAS \cite{supergradients} & \textbf{0.85} & \textbf{0.62} \\
RetinaNet \cite{wu2019detectron2} & 0.82 & 0.56 \\
Faster R-CNN \cite{wu2019detectron2} & 0.81 & 0.52 \\ \hline
\end{tabular}
}
\end{table}

\subsection{Quantitative Detection Evaluation}
\label{det:quant}
In this section, we show the results of the YOLO-NAS detector on the original and enhanced images of the CUPDD and RUOD. We call YOLO-NAS models that are trained on the Original images as \emph{original detectors}, and YOLO-NAS models that are trained on enhanced images as \emph{domain detectors}. On the one hand, the performance of the 10 YOLO-NAS original and domain detectors on the CUPDD dataset is shown in Table \ref{tab:CUPDD-DET}. The images of the CUPDD are severely degraded with very low visibility explaining the low overall performance of the original and the domain detectors, where the Original detector achieved the highest mAP by a notable margin at 0.38. The leading performance of the Original detector was evident in the Bushy class but not in the Leafy class where it outperformed by the AutoEnh domain detector by a notable margin of 3\%. This is possibly due to the fact that the bushy class has a more complex shape and edges that are obscured by enhancement noise compared to the simpler Leafy class. Unexpectedly, the PCDE achieves very close overall performance to the Original detector although we have discussed in section \ref{sec:enhancement} how the PCDE produces very low-quality and non-visually pleasing images. On the contrary, the ICSP has a severely deteriorated mAP performance at 0.22 aligning with the severe deterioration in its enhancement performance. The ACDC and TEBCF showed lower Leafy class mAP at 0.22 and have overall scores of 0.34 and 0.33, respectively. BayesRet showed consistently lower performance across all classes with an overall mAP of 0.30. AutoEnh achieved the highest performance in the Leafy class at 0.31 and an overall score of 0.37. Semi-UIR provided consistent detection score across all classes with an overall mAP of 0.34, matched by TUDA. Finally, USUIR stood out in the Tapey class at an mAP of 0.40 and achieved an overall mAP of 0.35. 

On the other hand, Table \ref{tab:RUOD-DET} shows the detection performance on RUOD dataset with similar results to Table \ref{tab:CUPDD-DET} where the Original detector again outperformed all other domain detectors, apart from the AutoEnh which had a matching overall performance. Other models achieved moderate performances, including ACDC, TEBCF, BayesRet, and USUIR that showed reasonably consistent mAP scores across most classes, resulting in overall mAP values between 0.59 and 0.61. Conversely, ICSP consistently ranked with the lowest scores across nearly all classes, yielding the lowest overall mAP at 0.55. There is a noticeable trend where all detectors perform best in the Diver and Cuttlefish classes, often achieving mAP scores above 0.70. Those two classes are mostly found in clear and high quality images and have relatively bigger instance size than other classes. An observable note is that the Jellyfish class achieved with the TUDA and BayesRet domain detectors around 3\% higher than the Original detector. These observations signify the variability in the detection performance using different enhancement methods and across different classes. In addition, no domain detector was able to outperform the original detector indicating a negative overall effect of enhancement on detection at the dataset level. However, this general conclusion might not hold true on an individual level where enhancement could have a positive effect on some images and a negative effect on others, resulting in a negative overall performance. Therefore, a more granular analysis that addresses the detection performance before and after enhancement at the image level is required.





\begin{table}
\centering
\caption{Per class detection evaluation of the original and enhanced CUPDD \cite{saleem2023multi} dataset by the selected enhancement models using the mean Average Precision (mAP$_{50-95}$).}
\label{tab:CUPDD-DET}
\resizebox{\linewidth}{!}{%
\begin{tabular}{>{\hspace{0pt}}m{0.558\linewidth}|>{\centering\hspace{0pt}}m{0.102\linewidth}>{\centering\hspace{0pt}}m{0.09\linewidth}>{\centering\arraybackslash}p{0.098\linewidth}>{\centering\arraybackslash\hspace{0pt}}m{0.083\linewidth}} \hline
\multicolumn{1}{>{\centering\hspace{0pt}}m{0.558\linewidth}|}{Methods} & Bushy & Leafy & Tapey & mAP \\ \hline
Original & \textbf{0.46} & 0.29 & \textbf{0.40} & \textbf{0.38} \\
ACDC \cite{ACDC} & 0.42 & 0.22 & 0.37 & 0.34 \\
TEBCF \cite{yuan2021tebcf} & 0.40 & 0.22 & 0.36 & 0.33 \\
BayesRet \cite{zhuang2021bayesian} & 0.30 & 0.23 & 0.36 & 0.30 \\
PCDE \cite{zhang2023underwater} & 0.43 & 0.30 & 0.39 & 0.37 \\
ICSP \cite{hou2023non} & 0.28 & 0.16 & 0.23 & 0.22 \\
AutoEnh \cite{tang2022autoenhancer} & 0.43 & \textbf{0.31} & 0.37 & 0.37 \\
Semi UIR \cite{huang2023contrastive} & 0.38 & 0.23 & \textbf{0.40} & 0.34 \\
USUIR \cite{fu2022unsupervised} & 0.40 & 0.25 & \textbf{0.40} & 0.35 \\
TUDA \cite{wang2023domain} & 0.38 & 0.27 & 0.38 & 0.34 \\ \hline
\end{tabular}
}
\end{table}

\begin{table*}[]
\caption{Per class detection evaluation of the original and enhanced RUOD \cite{fu2023rethinking} dataset by the selected enhancement models using the mean Average Precision (mAP$_{50-95}$).}
\label{tab:RUOD-DET}
\resizebox{\textwidth}{!}{%
\begin{tabular}{lccccccccccc}
\hline

\multicolumn{1}{l|}{Methods} &
  Holothurian &
  Echinus &
  Scallop &
  Starfish &
  Fish &
  Corals &
  Diver &
  Cuttlefish &
  Turtle &
  Jellyfish &
  mAP \\ \hline
\multicolumn{1}{l|}{Original} &
  \textbf{0.50} &
  \textbf{0.50} &
  \textbf{0.51} &
  \textbf{0.55} &
  \textbf{0.55} &
  \textbf{0.54} &
  \textbf{0.75} &
  \textbf{0.85} &
  \textbf{0.85} &
  0.59 &
  \textbf{0.62} \\
\multicolumn{1}{l|}{ACDC \cite{ACDC}}      & 0.48 & 0.48 & 0.48          & 0.53 & 0.52          & 0.51          & 0.73 & 0.83 & 0.84          & 0.60          & 0.60 \\
\multicolumn{1}{l|}{TEBCF \cite{yuan2021tebcf}}     & 0.48 & 0.49 & 0.49          & 0.53 & 0.52          & 0.49          & 0.71 & 0.81 & 0.81          & 0.58          & 0.59 \\
\multicolumn{1}{l|}{BayesRet \cite{zhuang2021bayesian}} & 0.48 & 0.47 & 0.49          & 0.53 & 0.53          & 0.52          & 0.72 & 0.83 & 0.84          & \textbf{0.62} & 0.60 \\
\multicolumn{1}{l|}{PCDE \cite{zhang2023underwater}}      & 0.48 & 0.49 & \textbf{0.51} & 0.53 & 0.54          & \textbf{0.54} & 0.73 & 0.84 & \textbf{0.85} & 0.59          & 0.61 \\
\multicolumn{1}{l|}{ICSP \cite{hou2023non}}      & 0.43 & 0.49 & 0.42          & 0.48 & 0.48          & 0.47          & 0.70 & 0.76 & 0.79          & 0.52          & 0.55 \\
\multicolumn{1}{l|}{AutoEnh \cite{tang2022autoenhancer}} &
  0.49 &
  \textbf{0.50} &
  \textbf{0.51} &
  0.54 &
  0.54 &
  \textbf{0.54} &
  0.74 &
  \textbf{0.85} &
  \textbf{0.85} &
  0.59 &
  \textbf{0.62} \\
\multicolumn{1}{l|}{Semi UIR \cite{huang2023contrastive}}  & 0.47 & 0.49 & 0.48          & 0.53 & \textbf{0.55} & 0.52          & 0.73 & 0.84 & 0.84          & 0.61          & 0.61 \\
\multicolumn{1}{l|}{USUIR \cite{fu2022unsupervised}}     & 0.47 & 0.49 & 0.49          & 0.53 & 0.54          & 0.53          & 0.73 & 0.84 & \textbf{0.85} & 0.60          & 0.61 \\
\multicolumn{1}{l|}{TUDA \cite{wang2023domain}}      & 0.47 & 0.48 & 0.47          & 0.53 & \textbf{0.55} & 0.52          & 0.73 & 0.84 & 0.84          & \textbf{0.62} & 0.61 \\ \hline
\end{tabular}%
}
\end{table*}

\subsection{Qualitative Detection Evaluation}
\begin{figure}[ht]
\centering
\includegraphics[width=\columnwidth]{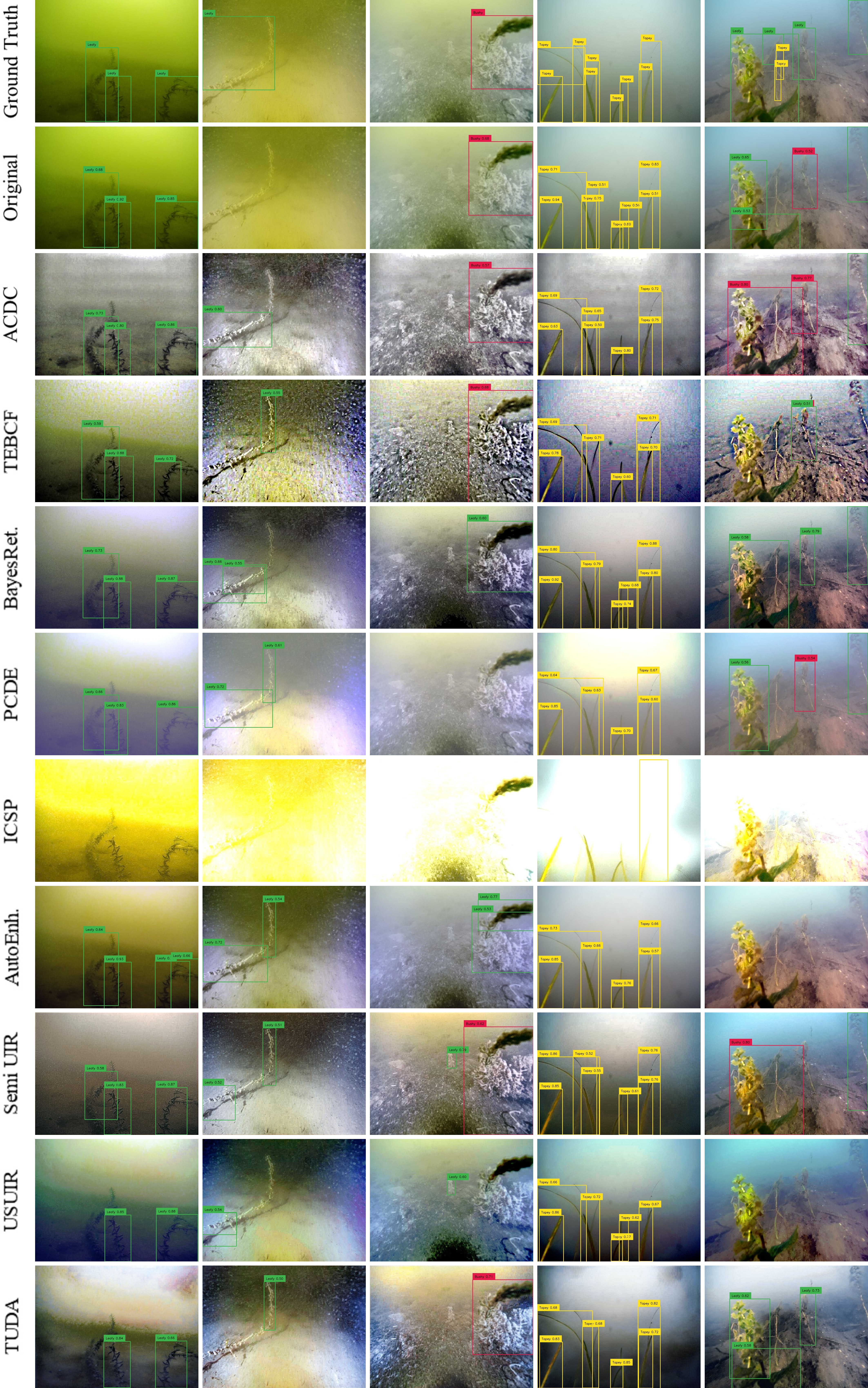}
  \caption{Inference visualization of the Original and domain detectors of five random images on the CUPDD dataset.}
  \label{det_vis_CUPDD}
\end{figure}

\begin{figure}[ht]
\centering
\includegraphics[width=\columnwidth]{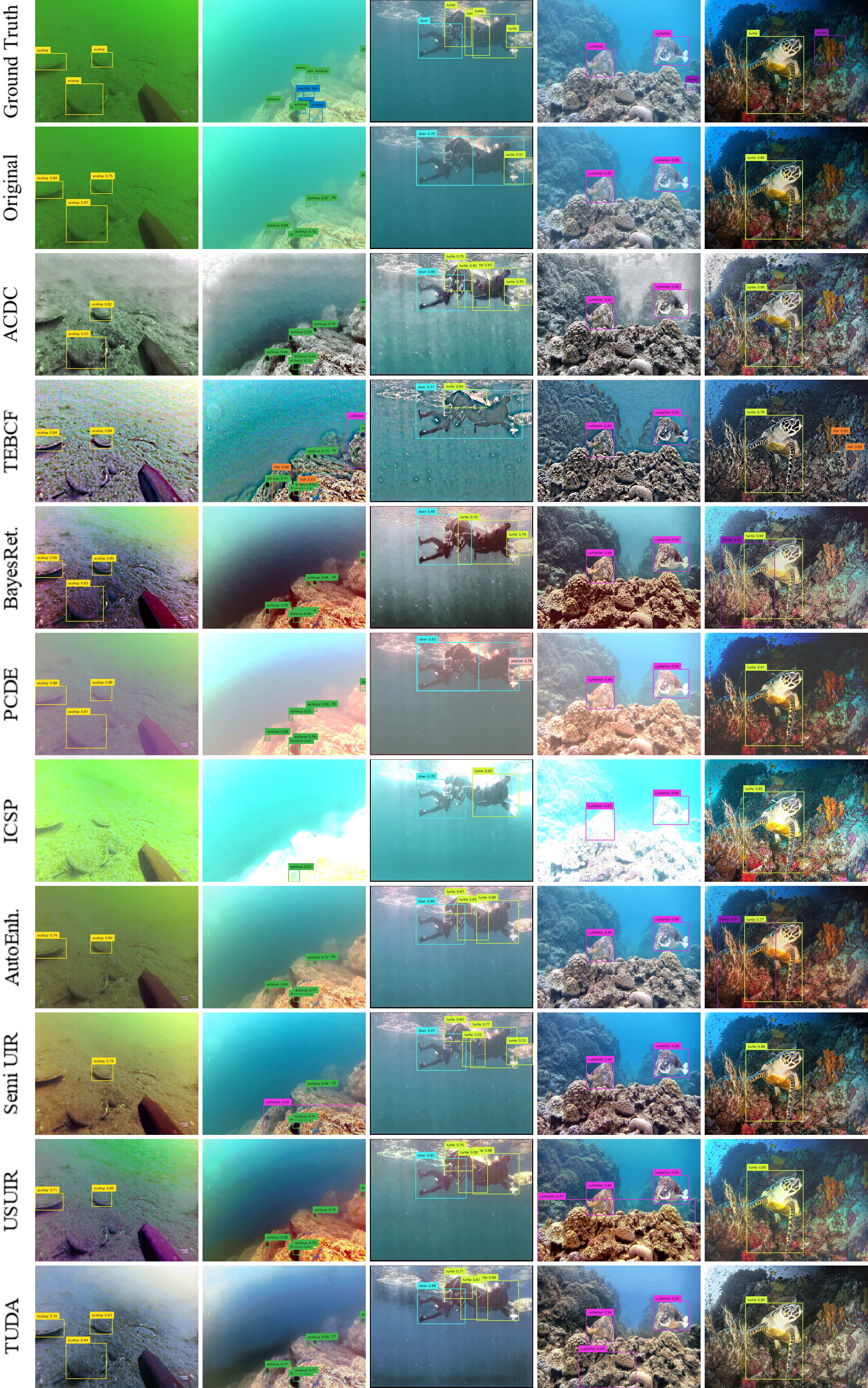}
  \caption{Inference visualization of the Original and domain detectors of five random images on the RUOD dataset.}
  \label{det_vis_RUOD}
\end{figure}

The inference from the original and domain detectors on five random testing images from both CUPDD and RUOD is provided Figs. \ref{det_vis_CUPDD} and \ref{det_vis_RUOD}. For a more convenient evaluation, we opt for a confidence threshold of 0.5. On the one hand, the Original detector achieved top performance detecting most objects correctly as shown in Fig. \ref{det_vis_CUPDD}, apart from the second and last images (counting from the left) where some objects were missed resulting in False Negatives (FNs) and other objects were confused resulting in False Positives (FPs). On the contrary, most domain detectors increased the FPs and FNs. In addition, the True Positives (TPs) detected by the domain detectors generally have lower Intersection over Union (IoU) thresholds compared with the Original detector which lowered the mAP$_{50-95}$ performance in section \ref{det:quant}. All domain detectors performed fairly well except for the PCDE and ICSP which have very foggy or overexposed images with little details and deformed objects' edges. We also observed that in the second image, the Original domain detector was not able to detect the Leafy plant but most domain detectors were able to detect at least parts of the Leafy plant. This indicates that the detection performance for this image increased after enhancement revealing the potential in enhancement improve the detection performance for some images. Furthermore, the Semi-UIR and the USUIR detected in the third image a leafy plant that is not in the ground truth. We think that human annotators could not see the Leafy plant due to the severe degradation of the original image. Therefore, this detected object will be considered a FP despite the fact that it should be a TP. 

On the other hand, similar trends are found in Fig. \ref{det_vis_RUOD} of the RUOD dataset where the Original detector outperformed domain detectors in most images. Although some models did not produce visually pleasing images, they were still able to achieve good performance such as the ACDC. In contrast, some models that did produce visually pleasing images underperformed other domain detectors. Similar to our observation on the CUPDD, we observed some cases where domain detectors outperformed the the Original detector, such as the case for the third image. We also observe that domain detectors that produce more noise and artifacts have a relatively higher number of FPs compared to other models, such as the case of the TEBCF model. We conclude that the overall negative impact of enhancement could be attributed to diffused edges and increased noise after enhancement. Furthermore, visually pleasing images do not necessarily lead to higher detection performance. In addition, image enhancement might improve the detection performance of some images and might also help in revealing hidden objects in the original images for human annotators.

\begin{figure*}[ht]
\centering
\includegraphics[width=\textwidth]{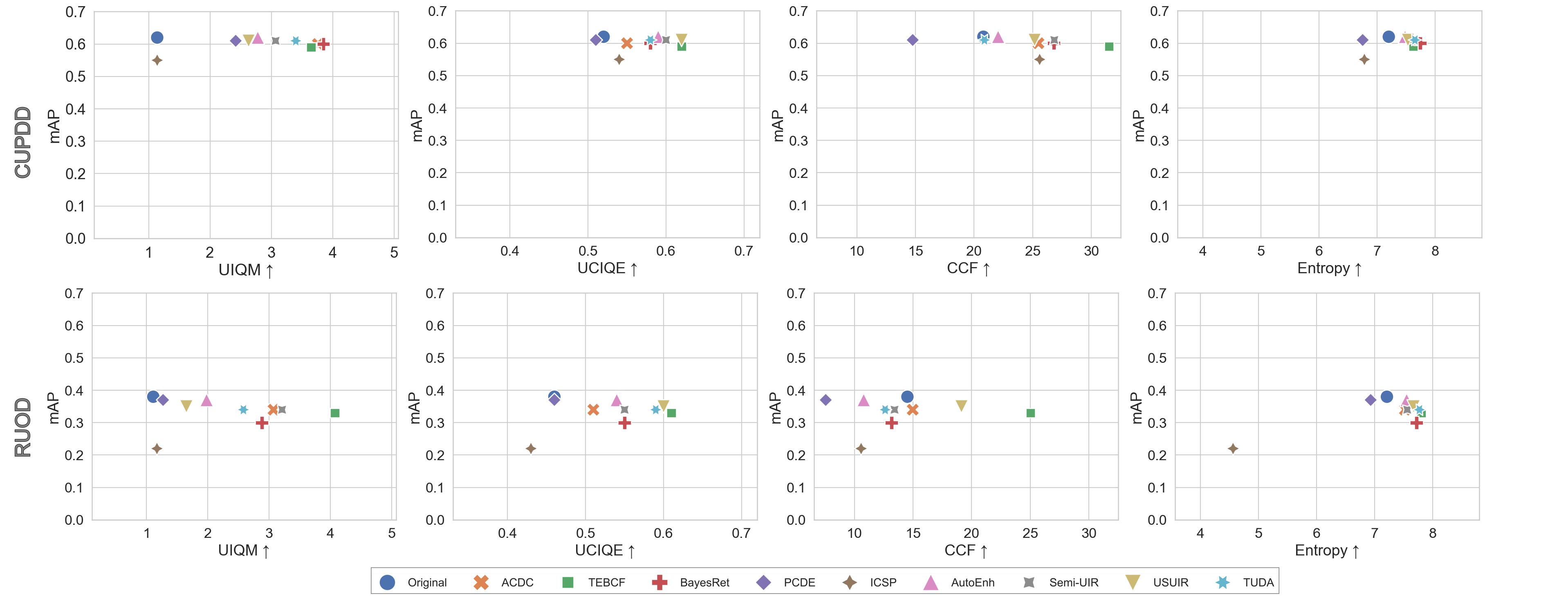}
  \caption{Scatter plots of the selected enhancement metrics against the mAP from the Original and domain detectors for CUPDD and RUOD datasets.}
  \label{corr}
\end{figure*}

\begin{figure}[t]
\centering
\includegraphics[width=0.75\columnwidth]{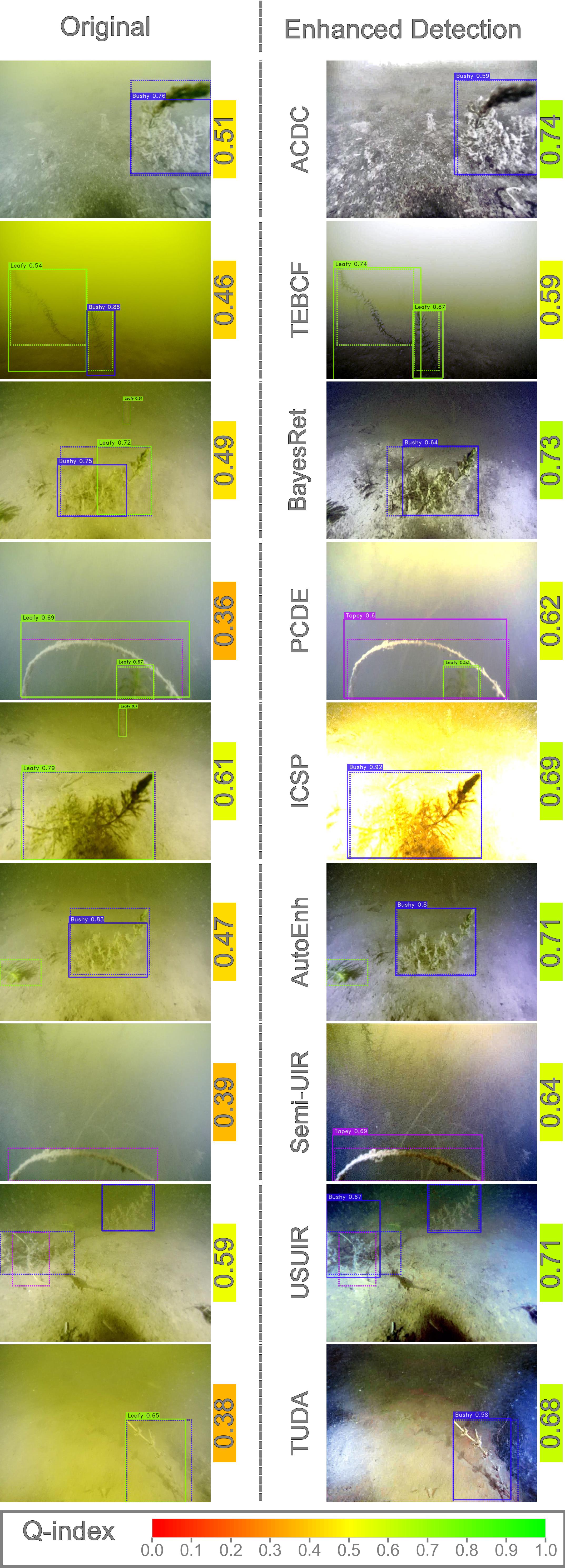}
  \caption{Cases where domain detectors performed better than the Original detector on CUPDD dataset. The ground truth bounding boxes are visualized on each image as dotted bounding boxes. The color-mapped values next to images represent Q-index}
  \label{CUPDD_ENH_DET}
\end{figure}

\begin{figure}[t]
\centering
\includegraphics[width=0.75\columnwidth]{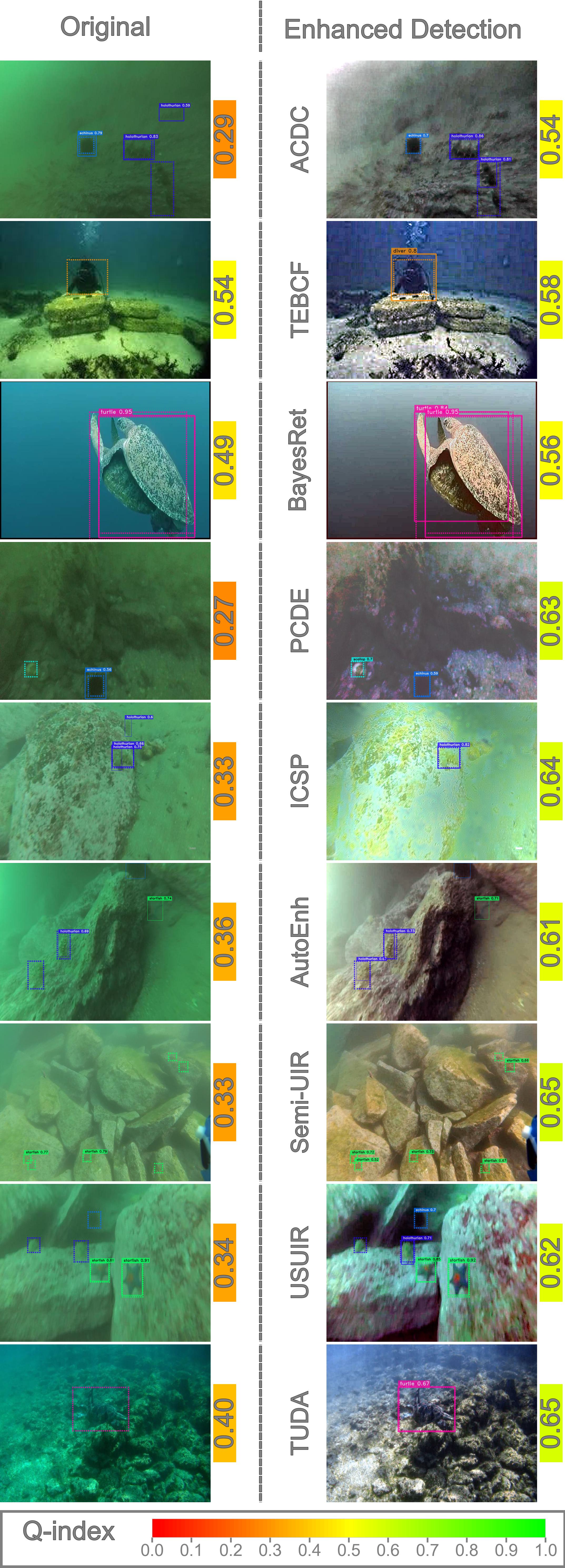}
  \caption{Cases where domain detectors performed better than the Original detector on RUOD dataset. The ground truth bounding boxes are visualized on each image as dotted bounding boxes. The color-mapped values next to images represent Q-index}
  \label{RUOD_ENH_DET}
\end{figure}

\section{Enhancement-Detection Discussion}
\label{sec:discussion}
In this section, we provide a combined enhancement-detection analysis. In particular, we study the correlation between the enhancement metrics and the mAP. In addition, we conduct a further image-level analysis and quantify cases where the detection performance improved after enhancement.

\subsection{Enhancement-Detection Metrics Correlation}
In this section, we generate scatter plots of the mean values of the selected enhancement metrics for each enhanced dataset and the mAP of the domain detectors for both the CUPDD and RUOD as shown in Fig.~\ref{corr}. The prevailing noticeable trend in Fig.~\ref{corr} across all metrics is that increases in enhancement metric values do not correlate with increases in mAP. For example, the mAP maintained similar values with the increase in the UIQM for the CUPDD, whereas it slightly decreased by the increase of the UIQM in the RUOD. Furthermore, some enhancement models, such as the AutoEnh and PCDE, achieved lower CCF scores on the RUOD compared to the original images, yet their images performed similarly in terms of the mAP. In contrast, the TEBCF and the USUIR achieved higher CCF but lower mAP compared to the original images. The ICSP model is an outlier on most plots and gained lower enhancement and mAP values due to its extremely overexposed images. The scatter plots presented in Fig.~\ref{corr} do not show any consistent relationship between quality and detection metrics with mixed and random performances across different models. This means that the enhancement performance does not reliably predict the detection performance. This discrepancy between the quality and detection metrics might be caused by the sensitivity of current quality metrics resulting in values that do not accurately reflect the true visual or structural quality of the images. Such outcomes may also be attributed to the fundamental differences between human and machine perception, further emphasizing the need for metrics that simultaneously account for visual quality from a human perspective and detection performance from a machine perspective.

\subsection{Image-Level Analysis}
In this section, we illustrate how enhancement improves detection by conducting a comprehensive qualitative evaluation, examining hundreds of images before and after enhancement to assess its effect on detection. Based on the insights gained from this analysis, we then perform a per-image quantitative analysis to validate the observed effects and derive more insights.

\subsubsection{Qualitative Analysis}
\label{Improving_Detection}
To further investigate the detection performance at the individual image level, we thoroughly surveyed individual images from both selected datasets and qualitatively compared their detection performance before and after enhancement. This time, we only look at cases where the domain detector performed better than the Original detector. This analysis is repeated for all domain detectors, and a representative image from each domain detector is selected to highlight the positive effect of enhancement on the detection performance as shown in Figs.~\ref{CUPDD_ENH_DET} and \ref{RUOD_ENH_DET}. To facilitate our comparison, the ground truth here is visualized using dotted boxes, whereas the detections are visualized using solid boxes. We first observed that the images presented in Figs.~\ref{CUPDD_ENH_DET} and \ref{RUOD_ENH_DET} are of low to medium quality, mostly achieving a Q-index below 0.5. Furthermore, as discussed in section \ref{q-dist} that enhancement improves the quality of low-quality images; all enhancement algorithms in Figs.~\ref{CUPDD_ENH_DET} and \ref{RUOD_ENH_DET} increased the Q-index of the original images. The lack of high-quality images in those cases, where the detection performance is better after enhancement, could indicate that enhancement of high-quality images results in lower detection performance compared to the original images. The enhancement and detection performances of different models varied in the CUPDD as shown in Fig. \ref{CUPDD_ENH_DET}. For example, the ACDC increased the contrast and removed the color cast of the Original image, resulting in an image that is not necessarily visually pleasing, but increased the detection performance with a much tighter box around the ground truth object. The TEBCF, BayesRet, and TUDA achieved a good enhancement performance with clearer objects' edges, turning some FPs into True Positives (TPs). Unexpectedly, the ICSP and PCDE also had cases where they enhanced the detection performance despite their overexposed or foggy images. The AutoEnh, Semi-UIR, and USUIR achieved similar performance and recognized undetected objects.

On the other hand, similar trends can be seen in Fig.~\ref{RUOD_ENH_DET} for RUOD dataset. It is worth noting that some models enhance one aspect of the image and deteriorate others, but still achieve better detection performance. For instance, the TEBCF removed the color cast of the image but introduced too much noise in the process, but was able to detect the diver object in the image anyway. This could be explained by the diver class being one of the easiest and most distinct classes to detect in RUOD dataset. The TUDA produced a very visually pleasing image, making it easier for its domain detector to distinguish the turtle from the underlying rocks. We conclude that enhancement has the potential to increase detection performance. This improvement may be due to the fact that the enhancement process reduces the disparity between low and high-quality images, as discussed in section \ref{q-dist}. i.e., By making low-quality, hard-to-detect testing scenes more similar to high-quality easy-to-detect training scenes within the same dataset, enhancement can result in better detection outcomes. These findings extend the conclusions made by previous studies \cite{fu2023rethinking, chen2020reveal, pei2019effects, wang2023underwater} that enhancement has a negative effect on detection, which is true at the dataset level, but not always true at the image level, as our evaluation suggests. Furthermore, most of the presented cases in Figs.~\ref{CUPDD_ENH_DET} and \ref{RUOD_ENH_DET} belong to low Q-index bins. This might indicate that exclusive enhancement for low-quality images could increase the detection performance compared to total enhancement of the entire dataset.

\begin{figure}[t]
\centering
\includegraphics[width=\columnwidth]{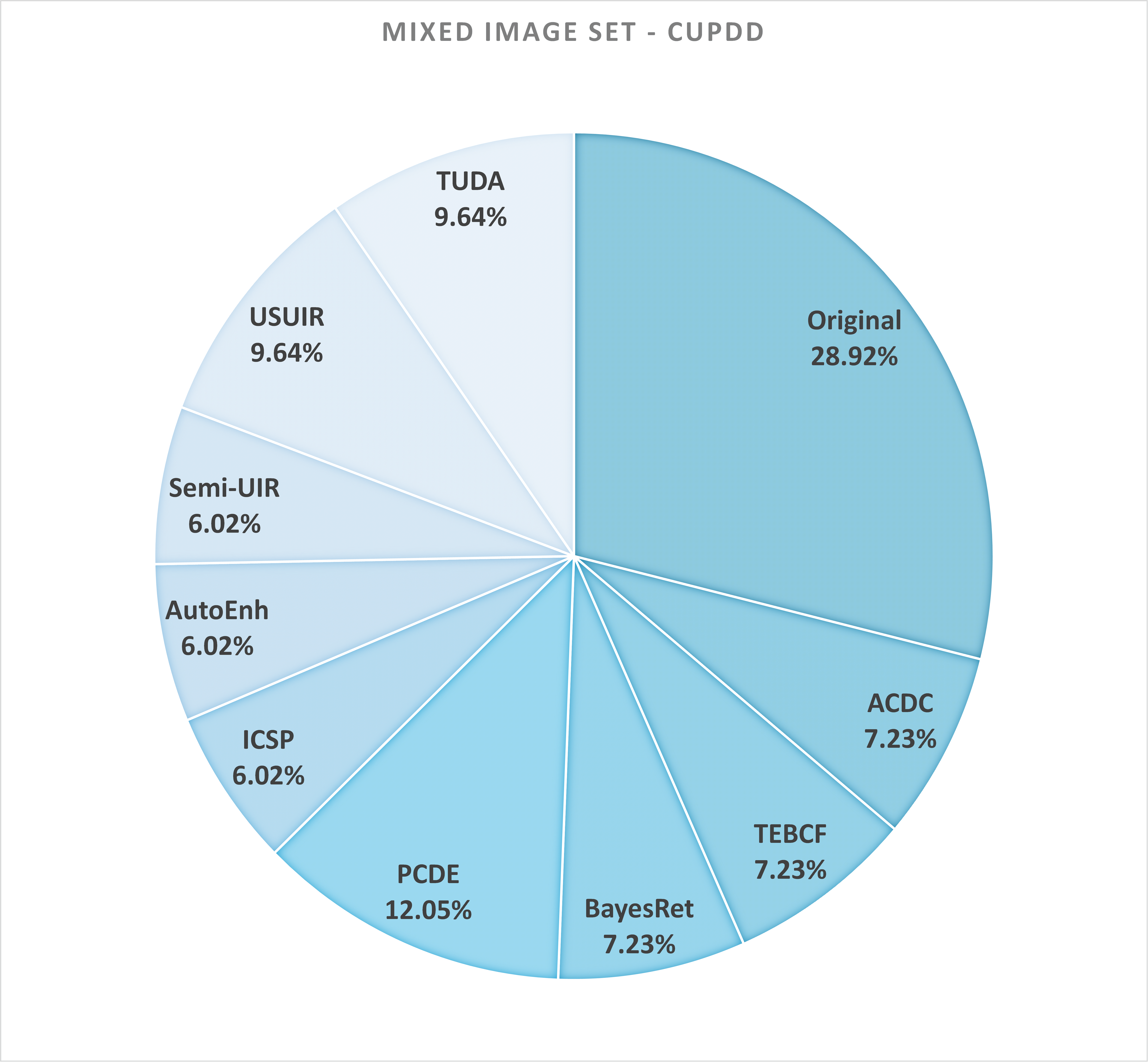}
  \caption{Pie chart of the generated mixed image set comprising original and enhanced images from the CUPDD test set \cite{saleem2023multi}. The images of this set are selected based on their mAP performance.}
  \label{CUPDD_pie}
\end{figure}

\begin{figure}[t]
\centering
\includegraphics[width=\columnwidth]{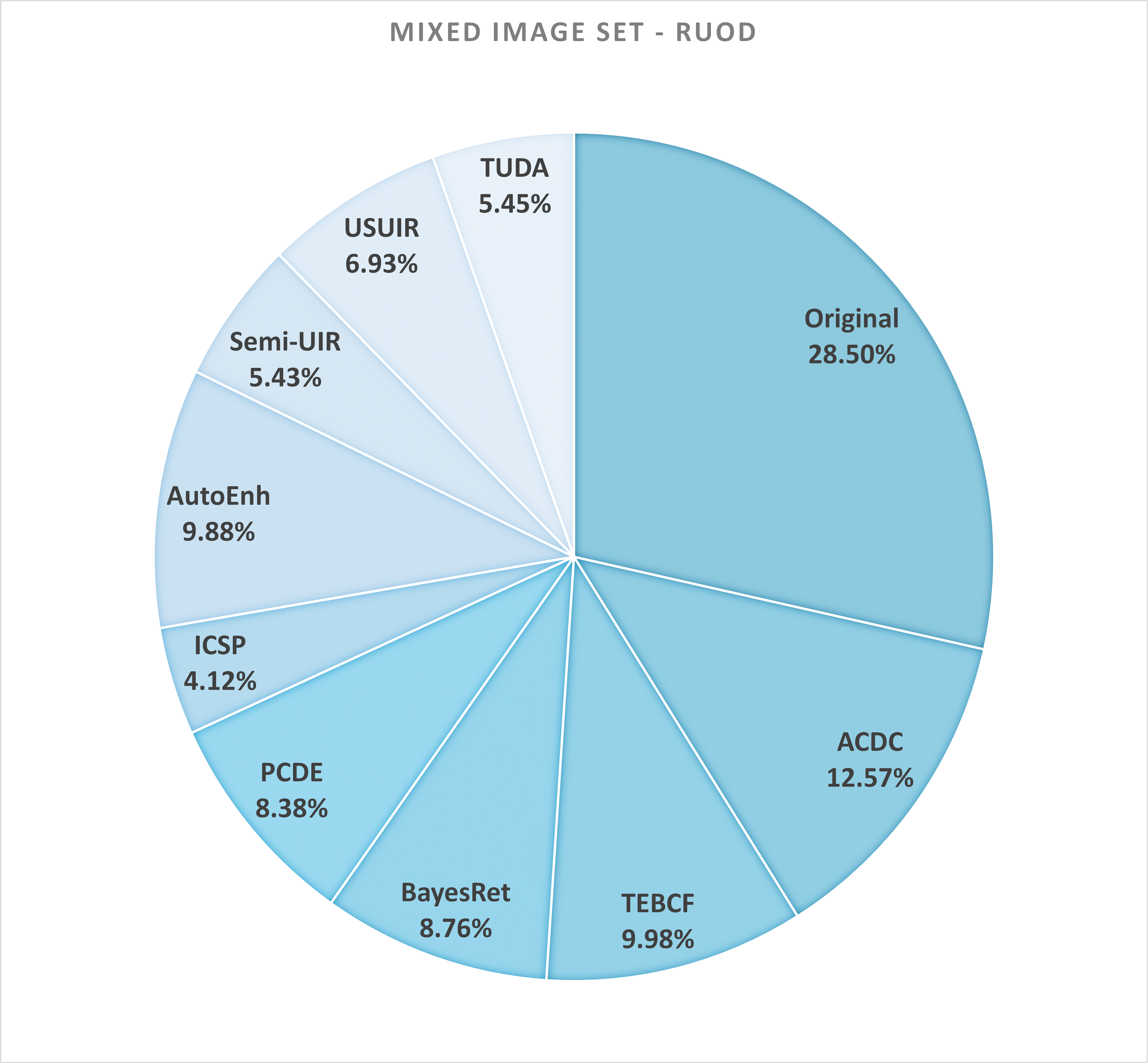}
  \caption{Pie chart of the generated mixed image set comprising original and enhanced images from the RUOD test set \cite{fu2023rethinking}. The images of this set are selected based on their mAP performance.}
  \label{RUOD_pie}
\end{figure}

\subsubsection{Quantitative Analysis}
Based on the observation that some images obtained higher detection performance after enhancement as discussed in Section \ref{Improving_Detection}, we conduct a further quantitative image-level evaluation on the original and all generated enhanced sets. In particular, the mAP is calculated for each original image and its corresponding enhanced versions. This allows us to compare the mAP performance across different variants of the same image. The variants associated with the highest mAP are selected to form a new mixed set, consisting of original and enhanced images produced by different enhancement algorithms. The per-image mAP values of the mixed set are then averaged to compute the overall mAP. Similarly, the per-image mAP values of all images in each individual set are averaged and presented in Table \ref{tab:mixed}. The mixed set achieves the highest possible mAP using the selected enhancement algorithms. Specifically, the mixed sets of CUPDD and RUOD achieved mAP scores of 0.64 and 0.77, reflecting improvements of 23\% and 9\%, respectively. The original sets recorded the second-best performance in both datasets, followed by different enhanced sets.

Furthermore, we analyze the contribution of each enhancement algorithm to the mixed set by calculating the proportion of images sourced from each algorithm. This is done by counting the number of images each enhancement algorithm contributed to the mixed set and dividing it by the total number of images in the test sets of CUPDD and RUOD, as illustrated in Figs.~\ref{CUPDD_pie} and \ref{RUOD_pie}. Both pie charts reveal similar trends, with the original images comprising the largest portion of the mixed sets. The remaining enhancement algorithms show relatively comparable contributions, ranging from approximately 12\% to 4.12\%.

This analysis highlights the potential of image enhancement to improve detection and support the development of further image-level evaluations aimed at selectively enhancing the images to boost detection performance without prior knowledge of their mAP performance.

\begin{table}
\centering
\caption{Per-image mean Average Precision (mAP$_{50-95}$) evaluation of the original, enhanced, and mixed image sets. The values shown represent the average of the individual mAP for each image in a set.}
\label{tab:mixed}
\resizebox{\linewidth}{!}{%
\begin{tblr}{
  width = \linewidth,
  colspec = {Q[63]Q[594]Q[144]Q[123]},
  row{1} = {c},
  row{2} = {c},
  row{3} = {c},
  row{13} = {c},
  row{14} = {c},
  cell{1}{1} = {c=2,r=2}{0.657\linewidth},
  cell{1}{3} = {c=2}{0.267\linewidth},
  cell{3}{1} = {c=2}{0.657\linewidth},
  cell{4}{1} = {r=9}{c},
  cell{4}{3} = {c},
  cell{4}{4} = {c},
  cell{5}{3} = {c},
  cell{5}{4} = {c},
  cell{6}{3} = {c},
  cell{6}{4} = {c},
  cell{7}{3} = {c},
  cell{7}{4} = {c},
  cell{8}{3} = {c},
  cell{8}{4} = {c},
  cell{9}{3} = {c},
  cell{9}{4} = {c},
  cell{10}{3} = {c},
  cell{10}{4} = {c},
  cell{11}{3} = {c},
  cell{11}{4} = {c},
  cell{12}{3} = {c},
  cell{12}{4} = {c},
  cell{13}{1} = {c=2}{0.657\linewidth},
  cell{14}{1} = {c=2}{0.657\linewidth},
  vline{2} = {1-14}{solid}, 
  vline{3} = {1-14}{solid}, 
  hline{1,3-4,13-15} = {-}{}, 
  hline{2} = {3-4}{},
}
\textbf{Set} &  & \textbf{Per-image mAP} & \\
 &  & CUPDD & RUOD\\
Original &  & 0.41 & 0.68\\
\rotatebox{90}{Enhanced} & ACDC \cite{ACDC} & 0.36 & 0.66\\
 & TEBCF \cite{yuan2021tebcf} & 0.35 & 0.66\\
 & BayesRet \cite{zhuang2021bayesian} & 0.35 & 0.66\\
 & PCDE \cite{zhang2023underwater} & 0.39 & 0.67\\
 & ICSP \cite{hou2023non} & 0.26 & 0.61\\
 & AutoEnh \cite{tang2022autoenhancer} & 0.37 & 0.68\\
 & Semi UIR \cite{huang2023contrastive} & 0.37 & 0.67\\
 & USUIR \cite{fu2022unsupervised} & 0.37 & 0.67\\
 & TUDA \cite{wang2023domain} & 0.38 & 0.67\\
\textbf{Mixed (CUPDD)} &  & \textbf{0.64} & \textbf{-}\\
\textbf{Mixed (RUOD)} &  & \textbf{-} & \textbf{0.77}
\end{tblr}
}
\end{table}

\section{Conclusion}
\label{sec:conclusion}
In this work, we conducted extensive experiments to evaluate recent underwater image enhancement methods and their effects on object detection. Our analysis included a joint quantitative and qualitative evaluation of different SOTA enhancement models on two datasets, namely RUOD and CUPDD. This joint analysis highlights a discrepancy between image quality metric values and human perception of quality, and reveals the limitations of current underwater metrics. In addition, a quality index (Q-index) was proposed to analyze the change in the quality distribution of the images after enhancement, revealing how enhancement improves the quality of low-quality images but over-enhances high-quality images, resulting in a negative quality effect. Moreover, we presented an evaluation of object detection performance on original and enhanced images, showing the negative effect of enhancement on detection at the dataset level. Furthermore, we conducted a correlation study between image quality and detection performance, showing no consistent relationship between image quality metrics and the mAP, and indicating that image quality cannot predict detection performance. Therefore, there is a pressing need to develop metrics that consider both image quality and detection at the same time. Finally, we conducted an image-level analysis instead of a dataset-level analysis to acquire conclusions at a granular level. In particular, we showcased images of improved detection performance after enhancement compared to the detection performance of the original images, indicating that enhancement has the potential to improve object detection. This is further validated by generating a top-performing mixed set containing original and enhanced images selected based on their mAP performance. The findings of this study provide insights for further investigation into the selective use of image enhancement on particular images to achieve improved overall detection performance.



\printcredits

\section*{Acknowledgment}
The authors appreciate the contributions of our colleagues and the technical assistance from the Great Lake Research Center and the United States Geological Survey (USGS). We extend our sincere appreciation to Anthony Geglio, Angus Galloway, Shadi Moradi, Phillipe A. Wernette, Alden T. Tilley, and Peter C.~Esselman who contributed their expertise and support throughout our study, making this manuscript possible.

\bibliographystyle{cas-model2-names}

\bibliography{cas-refs}


\bio{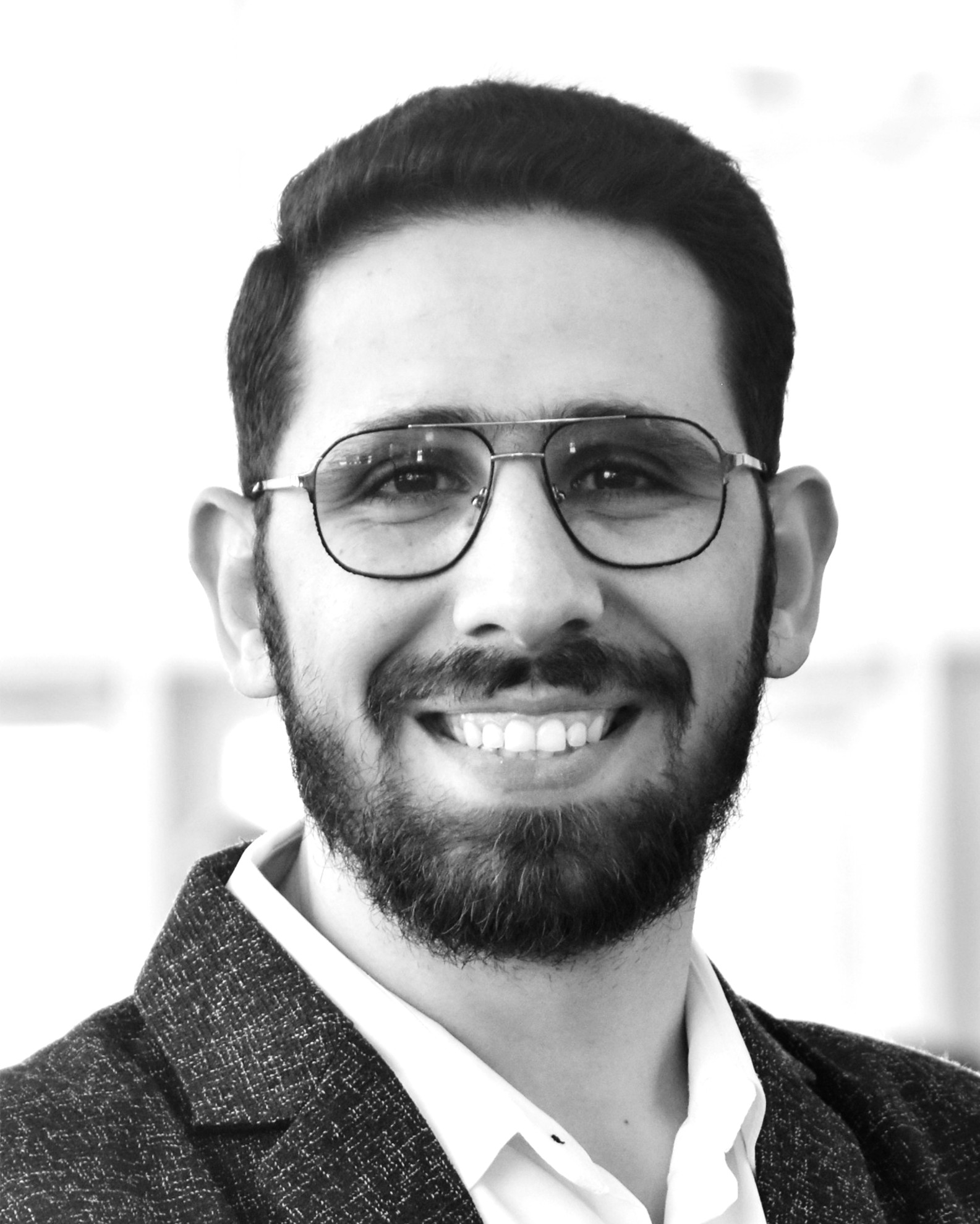}
Ali Awad is currently pursuing a Ph.D. in Computational Science and Engineering at Michigan Technological University (MTU). He earned his M.S. in Computer Engineering from the German-Jordanian University in 2019 and his B.S. in Computer Engineering from Philadelphia University, Jordan, in 2017. Over the past three years at MTU, Mr. Awad has developed a strong research background in computer vision, with a particular focus on underwater image enhancement and object detection. His work focuses on configuring image enhancement techniques to improve object detection performance, and he has published several papers in this area.
\endbio

\bio{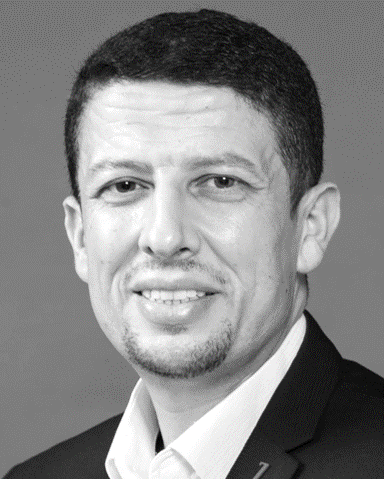}
Ashraf Saleem received his Ph.D. and MSc. degrees in mechatronics engineering from DeMontfort University, UK, in 2006 and 2003, respectively, and his BSc. in electrical and computer engineering from Philadelphia University, Jordan, in 2000. His research interests are unified under the theme, "deployment of robotics systems and artificial intelligence in the field of remote sensing", and his research focuses on solving real-life problems such as monitoring environmental pollution. He is also interested in developing real-time smart controllers for different engineering systems which include electromechanical, electro-pneumatic, and piezoelectric-based systems.
\endbio

\bio{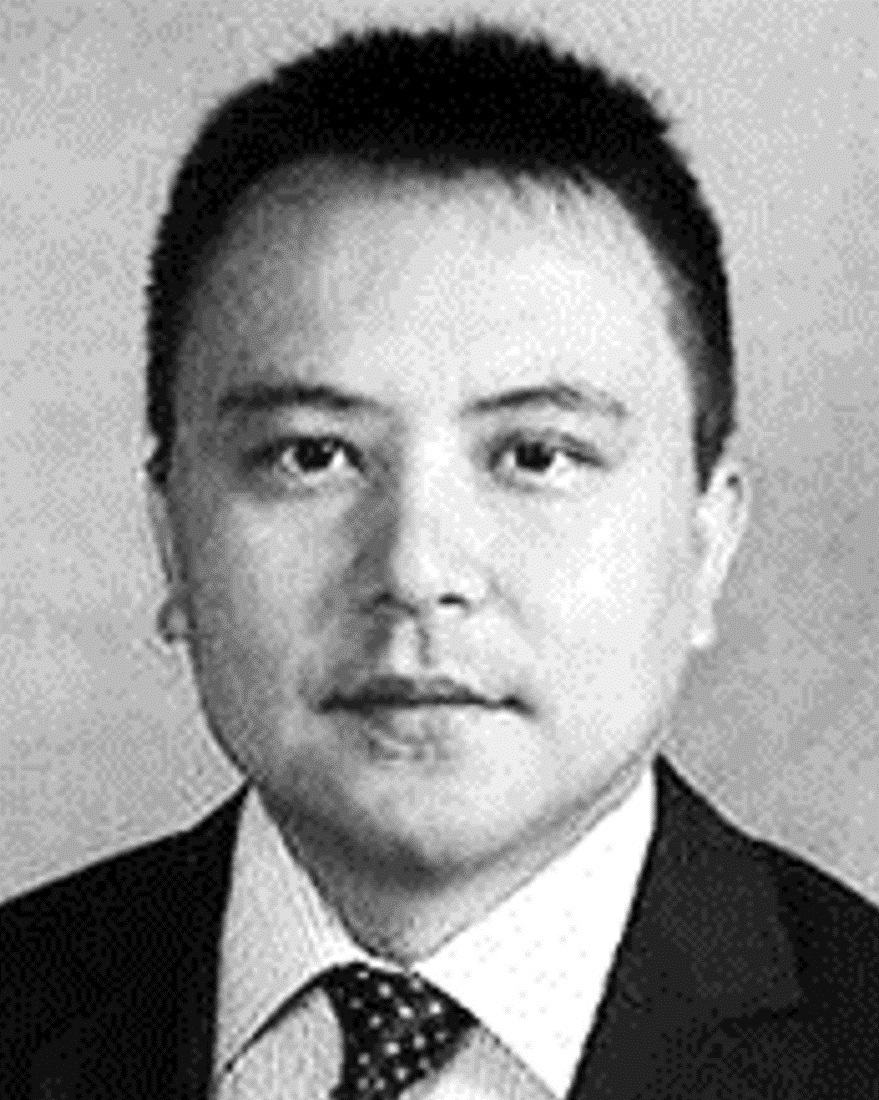}
Sidike Paheding received a Ph.D. in Electrical Engineering from the University of Dayton, Ohio, USA. Currently, he is an assistant professor in the Department of Computer Science and Engineering at Fairfield University. His research interests cover a variety of topics in image/video processing, machine learning, deep learning, computer vision, and remote sensing. Dr. Paheding is an associate editor of the Springer journal Signal, Image, and Video Processing, and ASPRS Journal Photogrammetric Engineering and Remote Sensing, and serves as a guest editor/reviewer for several reputed journals. Dr. Paheding is a Senior Member of the IEEE. 
\endbio

\newpage

\bio{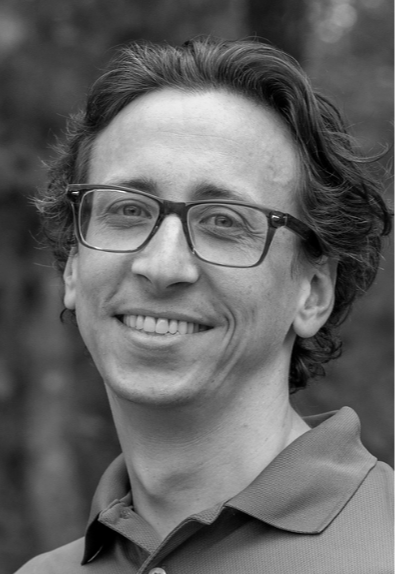}
Evan Lucas received his Ph.D. in electrical engineering from Michigan Technological University in 2023. His research interests are under the umbrella of applied machine learning and works in both the computer vision and natural language processing space. In computer vision, he is interested in underwater image enhancement and neural radiance fields methods for 3D scene reconstruction, the latter of which he co-founded a startup (Panverse Robotics) around. In natural language processing, he is currently interested in building educational chatbots and developing methods for detecting backdoor watermarks in generated text. 
\endbio

\bio{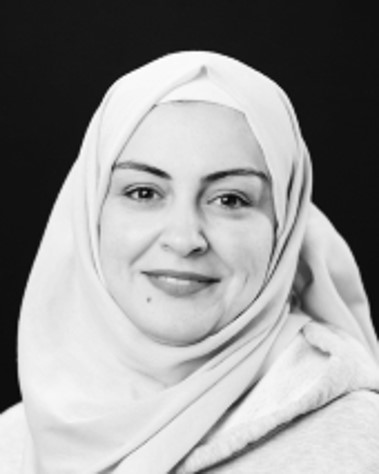}
Serein Al-Ratrout is an Assistant Teaching Professor with the Department of Computer Science at Michigan Technological University. She earned her PhD in Computer Science from De Montfort University, UK, in 2009. Dr. Al-Ratout's research interests focus on interactive learning, software development methodologies, optimization methods, and artificial intelligence. She has published numerous articles in these areas and conducted significant research projects. Notably, she led a study on heart rate variability to identify patients with obstructive sleep apnea. Dr. Al-Ratout is currently exploring the application of artificial intelligence in medical diagnosis, particularly through deep learning and neural networks.
\endbio

\bio{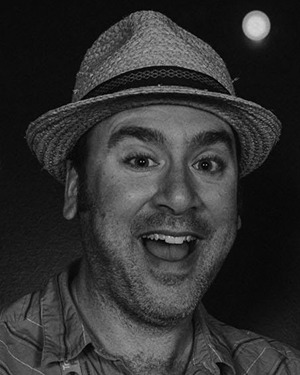}
Timothy C. Havens (Senior Member, IEEE) received B.S. and M.S. degrees in electrical engineering from Michigan Technological University and the Ph.D. degree in Electrical and Computer Engineering from the University of Missouri. Prior to joining Michigan Tech, he was an NSF/CRA Computing Innovation Postdoctoral Fellow at Michigan State University. Prior to his Ph.D. work, he was an Associate Technical Staff at MIT Lincoln Laboratory. He is the William and Gloria Jackson Associate Professor of Computer Systems and Associate Dean for Research in the College of Computing at Michigan Technological University. He is Director of the Institute of Computing and Cybersystems (ICC) and the ICC Center for Data Sciences. Dr. Havens has an active research program in the areas of sensor and data fusion, sensor and signal processing, and machine learning. He has published over 140 technical papers in various journals and conference proceedings. In 2012, Dr. Havens was awarded the Best Paper Award at the IEEE International Conference on Fuzzy Systems and in 2011 was awarded the IEEE Franklin V. Taylor Award for best paper at the IEEE Systems, Man, and Cybernetics conference. Dr. Havens is an Associate Editor for the IEEE Transactions on Fuzzy Systems and was a General Co-Chair of the 2019 IEEE International Conference on Fuzzy Systems.(Based on document published on 27 July 2020).
\endbio

\end{document}